\documentclass[twocolumn,letterpaper,amsmath,amssymb]{revtex4}
\usepackage{graphicx}
\usepackage{dcolumn}
\usepackage{bm}
\begin{document}

\title{Jamming III: Characterizing Randomness via the Entropy of Jammed Matter}

\author{Christopher Briscoe, Chaoming Song, Ping Wang, Hern\'an
A. Makse}

\affiliation {$^1$ Levich Institute and Physics Department, City
College of New York, New York, NY 10031, US}

\date{\today }

\begin{abstract}

The nature of randomness in disordered packings of frictional and
frictionless spheres is investigated using theory and simulations
of identical spherical grains.  The entropy of the packings is
defined through the force and volume ensemble of jammed matter and
shown difficult to calculate analytically.  A mesoscopic ensemble
of isostatic states is then utilized in an effort to predict the
entropy through the definition of a volume function dependant on
the coordination number. Equations of state are obtained relating
entropy, volume fraction and compactivity characterizing the
different states of jammed matter, and elucidating the phase
diagram for jammed granular matter. Analytical calculations are
compared to numerical simulations using volume fluctuation
analysis and graph theoretical methods, with reasonable agreement.
The entropy of the jammed system reveals that the random loose
packings are more disordered than random close packings, allowing
for an unambiguous interpretation of both limits. Ensemble
calculations show that the entropy vanishes at random close
packing (RCP), while numerical simulations show that a finite
entropy remains in the microscopic states at RCP. The notion of a
negative compactivity, that explores states with volume fractions
below those achievable by existing simulation protocols, is also
explored, expanding the equations of state. The mesoscopic theory
well reproduces the simulations results in shape, though a
difference in magnitude implies that the entire entropy of the
packing may not be captured by the herein presented methods.  We
discuss possible extensions to the present mesoscopic approach
describing packings from RLP to RCP to the ordered branch of the
equation of state in an effort to understand the entropy of jammed
matter in the full range of densities from RLP to FCC.
\end{abstract}

\maketitle

\section{Introduction}

Granular materials fall under the scope of athermal systems, which
includes glasses, colloids and gels, among others.  These athermal
systems exhibit non-equilibrium behavior, such that equilibrium
statistics is insufficient in its attempt to describe the system
dynamics.  These systems are thereby considered "complex", and
their characterization finds application in fields from chemistry
to fluid mechanics and beyond.  For granular systems, in
particular, a phase transition \cite{edwards,coniglio} occurs when
granular materials are compressed such that they develop a nonzero
stress in response to a strain deformation
\cite{liu-nagel,makse,zhang,ball,ohern}.  This transition,
referred to as the {\it jamming transition}, occurs at a critical
volume fraction, $\phi_c$, depending on interparticle friction and
preparation protocol. Analysis of the jamming transition produces
a phase diagram of jammed granular matter for identical spheres,
characterized by $\phi_c$ and the average mechanical coordination
number \cite{jamming2}.

The existence of boundaries in the phase diagram of
\cite{jamming2} are related to well-defined upper and lower limits
in the density of disordered packings; random close packing (RCP)
and random loose packing (RLP) \cite{bernal,onoda}.  How to
properly define RCP and RLP remains a longstanding open question
in the field.  It has been suggested \cite{edwards} that treating
a jammed system via the volume (V) ensemble introduces an analogue
to temperature in equilibrium systems.  This analogue,
"compactivity", is a measure of how compact a system could be.
Within this framework \cite{edwards,jamming2},  RCP is achieved in
the limit of minimal compactivity and RLP is achieved in the limit
of maximal compactivity.  Therefore, the boundaries of a phase
diagram for jammed matter could be defined by the limits of zero
and infinite compactivities.

\begin{figure}
\centering \resizebox{8.5cm}{!}{\includegraphics{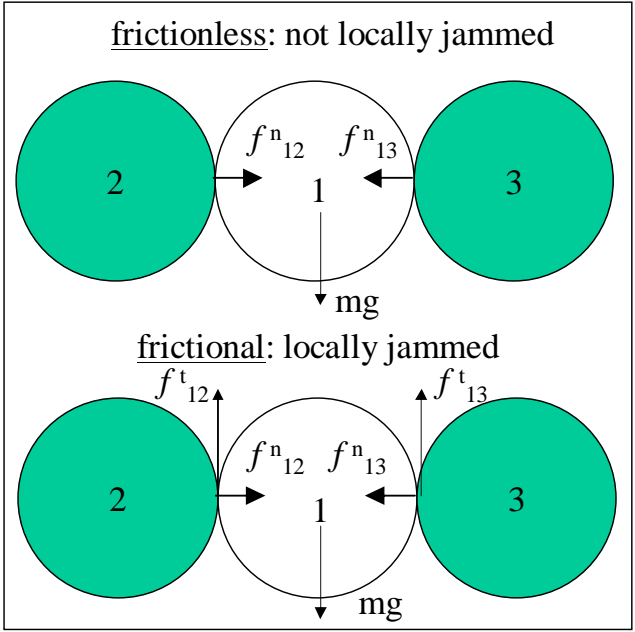}}
\caption{(a) The inset shows a ball in 2d under mechanical
equilibrium by two nearest neighbor contacts. The ball is not
jammed under a normal force interaction.  It jams when tangential
forces are present.}\label{sphere}
\end{figure}

In order to approach jammed systems with a statistical ensemble
approach, a definition of RCP and RLP requires proper definitions
of jammed states and the concept of randomness \cite{salvatore2}.
In an attempt to rigorously define jammed states, Torquato and
coworkers have proposed three categories of jamming
\cite{salvatore1}: locally, collectively and strictly jammed. This
definition is based purely on geometrical considerations and
therefore it is only sufficient for frictionless grains.
Frictional systems incorporate geometrical constraints but are
dominated by inter-particle normal and tangential contact forces
\cite{kertesz}.  In Fig. \ref{sphere} we see a hard sphere system
is not locally jammed if only normal forces are considered, since
the ball can freely move in the vertical direction.  The same
geometrical configuration is locally jammed if friction is allowed
between the particles, revealing the importance of forces in the
definition of jamming for frictional particles.  Therefore, a
definition of the jammed state for granular materials considering
only geometrical constraint is insufficient to describe frictional
grains.

Frictional systems further exhibit an inherent path dependency, as
granular contacts between grains result in the loss of energy
conservation. Approaches based on the potential energy landscape
\cite{ohern} thereby cannot be used for granular materials, as
such a potential does not exist for the non-conservative
frictional contact force. In this study, our framework is based on
statistical mechanics \cite{sirsam}, defining the jammed state at
the V-ensemble supplemented by force and torque balance
conditions, wherein volume replaces energy as the conservative
quantity for a statistical ensemble.  The free volume associated
with each particle in the packing is calculated as a function of
the geometrical coordination number using a coarse-grained
mesoscopic theory of quasi-particles.  This allows one to define
the RLP and RLP states through the ensemble of isostatic states.

Randomness in statistical systems is typically characterized by
the entropy, the equation of state derived from the number of
microstates available to the system.  In equilibrium statistical
mechanics, entropy provides the link between these microstates and
the macroscopic thermodynamic properties of the system.  We
explore that the concept of randomness is well-defined for the
V-ensemble following the Gibbs distribution \cite{sirsam}, and is
different from the measurement of randomness of single packing in
term of the ensemble of order parameters proposed in
\cite{salvatore2}. Therefore, calculating the entropy within the
V-ensemble can relate the available microscopic volume for each
grain to the macroscopic system properties, such as volume
fraction, average coordination number, and compactivity, in the
case of frictional hard spheres.

We first investigate frictional packings of equal sized spheres at
the jamming transition generated via computer simulations.  As the
volume fraction approaches the jamming transition from above,
$\phi\to\phi_c^+$, the system approaches the isostatic condition
and observables are shown to scale with the distance from the
jamming transition as a power law of $\phi-\phi_c$
\cite{makse,ball,ohern}, including stress, average coordination
number and elastic moduli.  We therefore consider the jamming
transition as the limit at which the stress tends to zero and the
average coordination number tends to a finite, non-zero, value.
Mechanical equilibrium imposes an average mechanical coordination
number, $Z$, larger or equal than the minimum isostatic
coordination as conjectured by Alexander \cite{alexander} (see
also
\cite{stealing,makse,zhang,ohern,kertesz,silbert,moukarzel,vanvan}).
The ensemble of packings at the jamming transition explores the
phase diagram of jammed matter by assuming the system is exactly
isostatic at the transition, and that the isostatic condition
varies as a function of the inter-particle friction coefficient
\cite{jamming2}.

We compute the equations of state, entropy and compactivity, as a
function of volume fraction, ranging from RLP to RCP.  The entropy
is calculated by two methods.  First, a direct analysis of volume
fluctuations via Einstein Fluctuation theory explores the
clustering of microscopic volumes. Second, graph theoretical
methods using the Shannon entropy analyze the network forming
properties of the granular system. These simulations reveal that
random loose packings have a higher disorder than random close
packings. Further, packings approaching RLP have a higher
compactivity than those approaching RCP.

Then we perform theoretical calculations under the quasi-particle
mesoscopic approximation of \cite{jamming2}, where a
coarse-graining over a mesoscopic length scale of several particle
diameters is implied, giving rise to a mesoscopic configurational
entropy, achieving a minimal value at the volume fraction of RCP
and maximal value at the RLP limit.  The results define RCP and
RLP at the mesoscopic level in general agreement with simulations,
suggesting that the concept of randomness in \cite{sirsam}
together with the notion of the jammed state in \cite{jamming2}
are useful.  The numerical results further suggest that the
mesoscopic entropy requires augmentation to include the entropy of
the microscopic states neglected at the mesoscopic level.

The maximal volume fraction for jammed spheres in 3d created using
purely random protocols occurs at $\phi_{\rm RCP} \simeq 0.64$.
Packings above RCP exist with some degree of order, up through the
perfectly ordered FCC state with $\phi_{\rm FCC} = 0.74$ in 3d. It
is of interest to understand how one would expand the existing
mesoscopic theory to include all packings from RLP to FCC, and
what effect this would have at the transition between disordered
and ordered packings at RCP.  While entropy tends to zero as we
approach FCC, the existing mesoscopic theory considers entropy
minimal at RCP, accounting only for disordered states.  It remains
an open topic as to whether a phase transition occurs at RCP,
noted by a discontinuous change in the equations of state, or
whether disorder decays smoothly when approaching FCC.  We discuss
propose plausible scenarios to rationalize the transition between
RCP and FCC as the volume fraction is increased by partial
crystallization. We speculate that at RCP a thermodynamic
transition, either continuous or discontinuous, may occur.  Such a
transition can be described by a full theory that includes both
ordered and disordered states and is beyond the scope of the
present work.  We stress that this is a hard sphere transition
different from the jamming transition obtained for deformable
particles as the external pressure approaches zero.  In this work,
hard sphere packings are numerically realized by simulating soft
particles in the limit of zero pressure using the "split"
algorithm explained in Section \ref{simresults}.A
\cite{makse,zhang,jamming2}.

Existing packing protocols exploring jammed packings may not probe
the entire phase diagram for jammed matter, as packings with a
negative compactivity may exist with volume fractions beneath the
minimum value \cite{rvlp}.  The mesoscopic theory is analyzed in
an effort to characterize these packings that are inaccessible via
random generation protocols.

\section{Simulations and Results}\label{simresults}

\subsection{Packing Preparation}

First, we investigate the entropy of jammed granular matter by
analyzing computer generated packings of 10,000 spherical
equal-size particles (a reader familiar with \cite{jamming2} can
refer to section II.C).

Two spherical grains in contact at positions $\vec{r}_1$ and
$\vec{r}_2$ and with radius $R$ interact with a Hertz normal
repulsive force \cite{landau} and Mindlin tangential contact
forces \cite{mindlin}.

The Hertz force is defined as:

\begin{equation}
F_n = \frac{2}{3}~ k_n R^{1/2}\delta ^{3/2},
\end{equation}
and an incremental Mindlin tangential force is defined as:

\begin{equation}
\Delta F_t= k_t (R \delta)^{1/2} \Delta s,
\end{equation}

Here the normal overlap is $\delta= (1/2)[2 R - |\vec{x}_1 -
\vec{x}_2|]>0$. The normal force acts only in compression, $F_n =
0$ when $\delta<0$.  The variable $s$ is defined such that the
relative shear displacement between the two grain centers is $2s$.
The prefactors $k_n=4 G / (1-\nu)$ and $k_t = 8 G / (2-\nu)$ are
defined in terms of the shear modulus $G$ and the Poisson's ratio
$\nu$ of the material from which the grains are made. We use
$G=29$ GPa and $\nu = 0.2$ typical values for spherical glass
beads and we use $R=5\times 10^{-5}$ m and the density of the
particles, $\rho=2 \times 10^{3}$ kg/m$^3$. Viscous dissipative
forces are added at the global level affecting the total velocity
of each particle through a term $-\gamma \dot{\vec{x}}$ in the
equation of motion, where $\gamma$ is the damping coefficient
related to the viscosity of the medium $\eta = \gamma / (6 \pi
R)$.  These dissipative forces ensure that the granular system
cannot 'rattle' forever.  We measure the time in units of
$t_0=R\sqrt{\rho/G}$, the compression rate in units of
$\Gamma_0=5.9 t_0^{-1}$ and the viscosity in units of $\eta_0= 8.2
R^2\rho/t_0$. The dynamics follows integration of Newton's
equations.

Sliding friction is also considered:
\begin{equation}
F_t \le \mu F_n.
\end{equation}
That is, when $F_t$ exceeds the Coulomb threshold, $\mu F_n$, the
grains slide and $F_t = \mu F_n$, where $\mu$ is the static
friction coefficient between the spheres.

The critical volume fraction at the jamming transition,
$\phi_{c},$ is achieved by the ``split'' algorithm as explained in
\cite{jamming2}, allowing one to obtain packings at the critical
density of jamming with arbitrary precision.

Initially, a dilute particle configuration is generated randomly,
usually with a volume fraction $0.30 \sim 0.36$.  Then, an
extremely slow isotropic compression, without friction, is applied
to this configuration until the system reaches $\phi_i$ in an
unjammed state.  For frictional packings the critical volume
fraction, $\phi_c$, is such that $\phi_c < 0.64$. Additionally,
the mechanical coordination number, $Z$, or the number of contacts
for a given particle which apply a force to maintain the jamming
condition at $\phi_c$, is $Z < 6$ for frictional systems.
Therefore, the system now at $\phi_i$ is allowed to relax while
maintaining the frictionless condition, such that the system is
unable to achieve jamming. $Z$ and the pressure decay to zero as
we relax the system below jamming.

After obtaining the relaxed, unjammed and frictionless state with
initial volume fraction $\phi_i$, a particular $\mu$ is given to
the particles and compression is applied with a compression rate
$\Gamma$ until a given volume fraction $\phi_1$. Then the
compression is stopped and the system is allowed to relax to
mechanical equilibrium by following Newton's equations without
further compression.  The split algorithm searches for $\phi_c$ by
setting upper and lower boundaries for $\phi_c$ and dividing the
size of those boundaries in half by iteratively compression (or
expanding) the system, followed by relaxing and then testing for a
non-zero stress in the system, as outlined in \cite{jamming2}.
This numerical process is repeated for packings with varying
$\phi_i$ along the range of $\mu$ between $0$ and $\infty$. The
results fill a phase diagram for jammed identical spheres right at
the jamming transition as obtained in \cite{jamming2} and shown in
Fig. \ref{phasesim}.

The friction coefficient ranges from 0 to $\infty$ producing
packings with coordination number varying from $Z\approx 6$ to
$Z\approx 4$, respectively. We find that there exits a common
function $Z(\mu)$ over the different $\Gamma$ and $\phi_i$ (see
\cite{jamming2}). For $\mu\to\infty$, $\phi$ ranges from the RLP
limit $\phi_{\rm RLP}\approx 0.55$ obtained when $\Gamma\to 0$ and
$\phi_i<0.55$ to the RCP limit $\phi_{\rm RCP}\approx 0.64$
obtained for larger $\Gamma$ and $\phi_i\to 0.64$. For $\mu=0$,
the density is approximately $\phi\approx\phi_{\rm RCP}$ with $Z
\simeq 6$.  All of the packings used herein, along with the
computer codes necessary to generate the packings and calculate
their entropy can be downloaded at http://jamlab.org.

\subsection{Phase Diagram}

Simple counting arguments, neglecting correlations between nearest
neighbors, consider that a necessary condition for mechanical
equilibrium is that the number of independent force variables must
be larger or equal than the number of linear independent
force/torque balance equations.  Alexander \cite{alexander}
conjectured that at the transition point for frictionless
spherical packings \cite{alexander,stealing,ball} the system is
exactly isostatic with a minimal coordination, $Z = 2d=6$ in 3d.
Such a conjecture can be extended to the infinite friction case,
where $Z = d+1=4$ \cite{edwards,ball}.  In the presence of finite
inter-particle friction coefficient $\mu$, there exists a
dependency of $Z$ and $\mu$ suggested by simulations
\cite{kertesz,jamming2,silbert}.

\begin{figure}
\centering { \vbox{
 \resizebox{8cm}{!}{\includegraphics{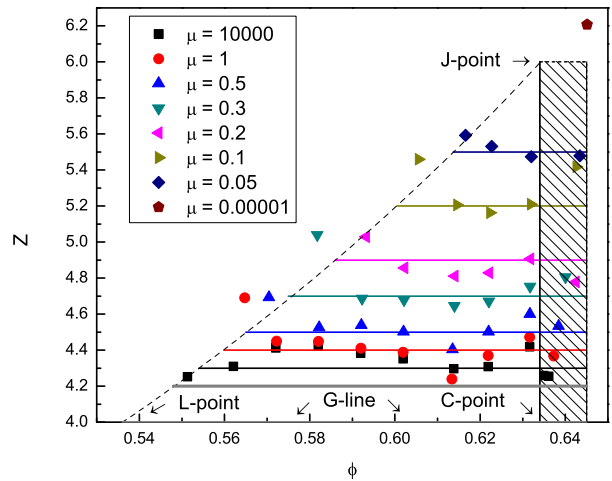}}
} } \caption{The phase diagram of jamming from simulation results.
We use the same algorithm to generate the packings as done in
\cite{jamming2}.  Here, we use 10,000 particles, while in
\cite{jamming2} only 1000 were used.  A larger number of particles
is necessary for accurate entropy calculations.  The volume
fraction is denoted by $\phi$ as opposed to $\phi_c$ for
simplicity.  Horizontal lines show the average coordination number
used for packings of constant $\mu$. The dashed line represents
the theoretical RLP line. The solid vertical line at $\phi =
0.634$ is the theoretical RCP line obtained in \cite{jamming2}.
Notice that some packings exist to the right of the RCP line. Such
packings are not captured by the theory, indicating that
microscopic fluctuations beyond the mesoscopic theory of
\cite{jamming2} are important close to the RCP state. The solid
grey line at $Z = 4.2$ indicates the lower limit for $Z$ available
using the present split algorithm.  The J-point, located at
($\phi, Z) = ($0.634,6), is the theoretical frictionless jamming
point.  The L-point, located at ($\phi, Z) = ($0.536,4), is the
theoretical jamming point for $\mu \rightarrow \infty$ with $X
\rightarrow \infty$. The C-point, located at ($\phi, Z) =
($0.634,4.0), is the theoretical jamming point for $\mu
\rightarrow \infty$ with $X \rightarrow 0$.  The G-line, $Z = 4.0$
is the theoretical average $Z$ achieved for all infinite friction
packings of identical spherical grains in 3d. } \label{phasesim}
\end{figure}

Figure \ref{phasesim} shows the phase diagram used for all
equation of state calculations presented herein, as obtained using
the "split" algorithm as described above in the plane $(\phi_c,
Z)$ (for simplicity, in what follows we denote $\phi = \phi_c$.
That is, we understand that all packings considered herein are at
the jamming transition and are hard sphere packings). As discussed
in \cite{jamming2}, packings along the RCP line for finite $\mu$
are most difficult to obtain, most notably near C point in Fig
\ref{phasesim}, resulting in higher values of the lowest
achievable stress for those packings. The G-line, at $Z = 4.0$,
indicates the theoretical $Z$ for infinite friction packings.  The
grey line at $Z = 4.2$ indicates the approximate lowest achievable
$Z$ possible using the present "split algorithm". The solid color
lines in Fig. \ref{phasesim} are averages used in all following
calculations.  Near the RLP and RCP lines for a some fixed $\mu$,
we observe notably higher values of $Z$.  These values are not
included in the average. This allows us to use a constant value as
an approximation for the mechanical coordination number.

We compare this result to the phase diagram as predicted by
theoretical model asserted in \cite{jamming2}.  Note that the
isostatic condition \cite{alexander} predicts $Z = 6$, while Fig.
\ref{phasesim} includes packings with $6 < Z \leq 6.2$, and $\phi
> 0.634$ as predicted by the theory for RCP.  We suggest that these packings are new microstates of jammed
matter (indicated by the shaded portion of the phase diagram)
which are not accounted for in the mean-field version of
\cite{jamming1, jamming2}. While they remain a component of the
ensemble generated via the above described simulation protocol,
their existence remains a topic of ongoing study.

\subsection{Entropy from Voronoi Volume Fluctuations}

In the absence of energy conservation, a different statistical
approach is necessary to describe the ensemble properties of
jammed granular matter. Along this line of research, Edwards
\cite{sirsam} proposes replacing the system energy by the volume
as the conservative quantity such that a microcanonical partition
function of jammed states can be defined and a statistical
mechanical analysis is plausible.  Therefore, a microscopic volume
must be associated with each grain.

As detailed in Jamming I \cite{jamming1}, the definition of a
Voronoi cell is a convex polygon whose interior consists of all
points closer to a given particle than to any other.  Further, it
is additive and tiles the system volume completely.  The formula
for the Voronoi volume of a particle, $i$, in terms of particle
positions for monodisperse spherical packings in 3d is
\cite{jamming1}
\begin{equation}
  \label{vor1}  {\cal W}_i^{\rm vor} = \frac{1}{3}\int
  \left(\frac{1}{2R}\min_{j}\frac{r_{ij}}{\cos \theta_{ij}}\right)^3
  ds,
\end{equation}
where $\vec{r}_{ij}$ is the vector from the position of particle
$i$ to that of particle $j$, the integrand is over all the
directions $\hat{s}$ forming an angle $\theta_{ij}$ with
$\vec{r}_{ij}$, and $R$ is the radius of the grain.  The Voronoi
volume is used to tile the total system volume, and replaces
energy as the conserved quantity in a new micro-canonical ensemble
for jammed granular matter. Therefore, fluctuations in Voronoi
cell volumes are related to the compactivity of the jammed system,
much like energy fluctuations are directly related to the system
temperature in equilibrium thermodynamics.  We notice that the
Voronoi-Delaunay decomposition is the basis for Hales proof of the
Kepler conjecture \cite{Hales}.  Below, we treat the monodisperse
case.  Other cases will be treated in subsequent papers.

\begin{figure}
\centering {
\vbox{\resizebox{8cm}{!}{\includegraphics{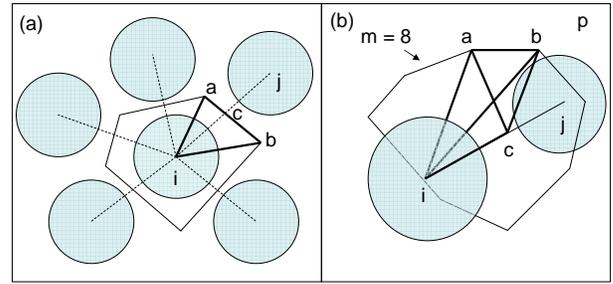}} } }
\caption{(a) Example of 2d Voronoi cell - All of the calculations
are done in 3d but are shown here in 2d for simplicity.  The line
between the centers of particles $i$ and $j$ is defined by
$\overline{ij}$, equivalent to $r_{ij}$. The line perpendicular to
the bisection of $\overline{ij}$ is defined by $\overline{ab}$,
intersecting $\overline{ij}$ at point $c$. $4$ additional
particles are Delaunay contacts of particle $i$, such that a $5$
sided polygon (pentagon) surrounds particle $i$, defining the
Voronoi cell of particle $i$ by virtue of intersecting bisecting
lines between each pair of Delaunay contacts.  Points $a$ and $b$
are defined as the boundary of the Voronoi cell line between
particles $i$ and $j$.  A triangle is thereby formed by points $i$
and $\overline{ab}$, the area of which is the contribution of the
Voronoi cell of particle $i$ exclusive with its Delaunay contact
to particle $j$.  This process is repeated for all Delaunay
contacts of $i$ to give the entire Voronoi cell area.  Note that a
symmetric area to $\triangle iab$ exists as $\triangle jab$, and
can be applied to the Voronoi cell area of particle $j$.  (b)
Example of 3d Voronoi cell - The line between the centers of
particles $i$ and $j$ is defined by $\overline{ij}$, equivalent to
$r_{ij}$. The plane perpendicular to the bisection of
$\overline{ij}$ is defined by $p$, intersecting $\overline{ij}$ at
point $c$.  Note that particles $i$ and $j$ are identical spheres,
with particle $j$ appearing smaller only to illustrate the 3d
properties of the system.  Plane $p$ is intersected by $m$ other
planes, creating an $m$-sided polygon between particles $i$ and
$j$.  Each plane intersecting plane $p$ (not shown) is a plane
bisecting $\overline{ik}$, the line between the centers of
particle $i$ (or $j$) and another particle $k$ in the system,
where $k$ is one of $m$ particular particles. A pyramid is thereby
formed using the $m$-sided polygon as the base, and $i$ (or $j$)
as the apex.  This pyramid is symmetric over plane $p$, and its
volume is the contribution to the Voronoi volume of particle $i$
from particle $j$, or vice versa, exclusively.  The volume of the
pyramid is calculated by separating the pyramid into $8$ smaller
pyramids, using the triangle composed of one of the $m$ available
sides, and $c$ as its base and $i$ as its apex. This is
illustrated by using $\overline{ab}$ and $c$ as the base of a
pyramid with apex $i$. The volume of this pyramid is calculated
and the process is repeated for each of the $m$ sides, adding each
obtained volume to the Voronoi volume of both $i$ and $j$.  The
entire process is then repeated for all Delaunay contacts for a
given particle, resulting in the total Voronoi volume for that
particle.  } \label{polygon}
\end{figure}

Next, we calculate the entropy of the numerical packings in Fig.
\ref{phasesim} from Voronoi volume fluctuation analogous to
Einstein Fluctuation theory. We first define the Voronoi cell
associated with each particle $i$ and calculate its Voronoi volume
${\cal W}_i$. Calculation of a Voronoi cell volume begins by
defining the polygon between two Delaunay contacts, having finite
number, $m$, vertices. Two grains are considered Delaunay contacts
if their corresponding Voronoi cells share a face. Delaunay
contacts are determined by the network of grain positions and
radii calculated using QHull software, available at
http://www.qhull.org. The contribution of this polygon to
calculating the Voronoi volume comes from the ability to associate
a pyramid, comprised of the center of each particle as the apex,
and the m-sided polygon as its base, as shown in Fig.
\ref{polygon} (schematically in 2d for simplicity), to each
particle. The two pyramids are symmetric. The volume of this
pyramid is the contribution to the Voronoi volume of the cell
surrounding a particle, exclusive to the particular Delaunay
contact which shares the polygon base. Repeating this process for
each Delaunay contact results in the complete Voronoi volume
surrounding a particle.  The Voronoi volume is thereby the
microscopic volume associated with each grain.

We perform statistical analysis of the volume fluctuations by
considering a cluster of $n$ particles. The Einstein fluctuation
relation is defined as follows \cite{chicago2,swinney}:

\begin{equation}
\sigma_{n}^2 \equiv \langle ({\cal W}_n - \langle {\cal
W}_n\rangle )^2 \rangle = \lambda X^2 d\langle {\cal W}_n\rangle /
dX \label{compac1}
\end{equation}

Equation (\ref{compac1}) is analogous to equilibrium
thermodynamics, replacing energy and temperature by volume and
compactivity, $X$, in the Edwards picture.  Note that $\lambda$ is
the analogue of the Boltzmann constant $k_B$ that defines the
units of compactivity.

\begin{figure}
\centering { \vbox{
 \resizebox{8cm}{!}{\includegraphics{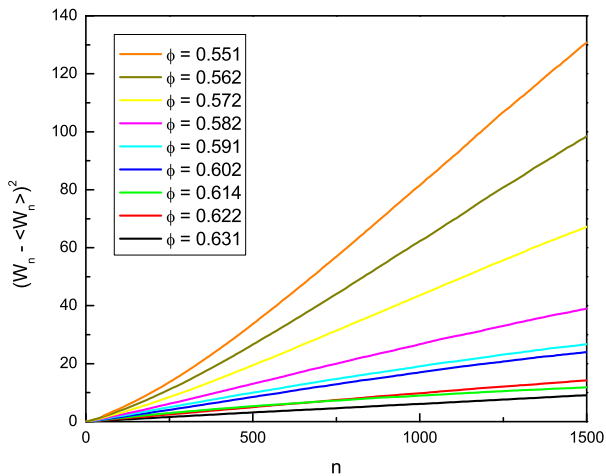}}
} } \caption{$\langle \Delta {\cal W}_n^2\rangle $ versus $n_c$
for packings with $\mu \rightarrow \infty$, with $Z \approx 4.3$,
for different $\phi$.} \label{fluctuations}
\end{figure}

We calculate the average volume, $\langle {\cal W}_n\rangle$ and
fluctuations $\sigma_n \equiv \langle ({\cal W}_n - \langle {\cal
W}_n\rangle )^2 \rangle$, where $\langle \cdot \rangle$ is an
average over many $n$-clusters.  From the large $n$ behavior we
extract the fluctuations versus volume fraction, $\phi$, for every
packing depicted in Fig \ref{phasesim}. Figure \ref{fluctuations}
shows the fluctuations as a function of $n$ for packings with
infinite friction, displaying the largest range of volume
fractions in the data used herein.  We find that for sufficiently
large $n \gg n_c$, the fluctuations scale with $n$ and therefore
are extensive and well-defined.

\begin{figure}
\centering { \vbox{
 \resizebox{8cm}{!}{\includegraphics{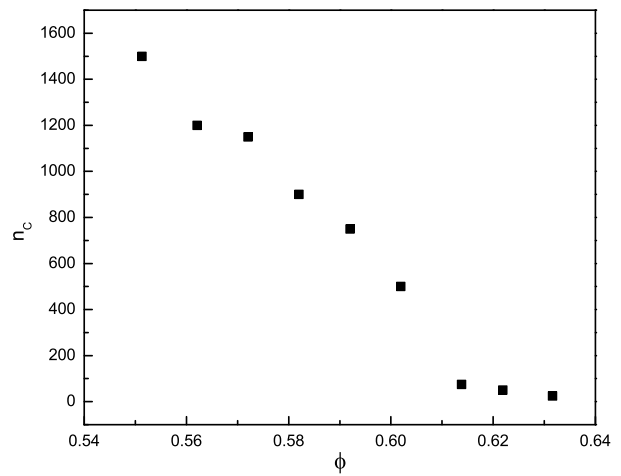}}
} } \caption{$n_c$ vs $\phi$ showing maximal value of extensive
nature.  Values of $\phi$ are taken along $\mu \rightarrow
\infty$, with $Z \approx 4.3$, to display largest range of $\phi$.
} \label{extensive}
\end{figure}

\begin{figure}
\centering { \vbox{
 \resizebox{8cm}{!}{\includegraphics{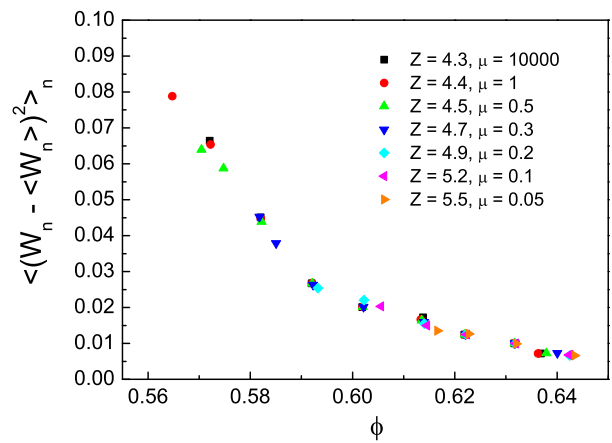}}
} } \caption{$\langle \Delta {\cal W}_n^2\rangle_n $ versus $\phi$
for packings with different friction coefficients producing
different mechanical coordination number versus the volume
fraction.  Each dataset with fixed coordination number corresponds
to the packings along each horizontal line in Fig. \ref{phasesim}.
} \label{fluctuationdens}
\end{figure}

Figure \ref{extensive} shows the approximate value of $n_c$ at
which the extensive nature of the fluctuations reaches its maximal
value, as a function of $\phi$. Lower volume fractions,
approaching RLP, require higher values of $n \approx 1000$ to
achieve this condition.  This result contrasts with the results of
\cite{dauchot}, although there the system was smaller, $N\approx
100$, and two-dimensional.  Reference \cite{dauchot} acknowledges
that if grain volumes can be treated as independent random
variables, then the fluctuation in clusters of $n$ Voronoi cell
volumes should scale with $n$. For clusters of jammed grains, this
was not observed in \cite{dauchot}, indicating the existence of
correlations between the Voronoi cell volumes within a cluster.
However, in the present study, the value of $n$ at which the
fluctuations are extensive is a function of $\phi$, shown in Fig.
\ref{extensive}. In \cite{dauchot}, this phenomena is observed,
but the density is apparently independent of the volume fraction,
and occurs at the same value of $n$ for each packing. For packings
approaching RCP, the extensive nature of the fluctuations occurs
at $n_c\approx 100$, lower than $n_c \approx 1000$ for RLP. Figure
\ref{fluctuationdens} shows the fluctuation density, $\langle
({\cal W}_n - \langle {\cal W}_n\rangle )^2 \rangle_n =
\frac{\langle ({\cal W}_n - \langle {\cal W}_n\rangle )^2
\rangle}{n}$, or fluctuation per grain, as a function of $\phi$
for all packings used herein obtained for $n > n_c$.  This data
represents the main equation of state for all of the numerical
packings considered herein.  Each color curve represents packings
with a fixed $Z$ (or $\mu$) as indicated in the figure.  We note
that while the fluctuations for all $Z(\mu)$ in this study
collapse onto a single curve, as illustrated in Fig
\ref{fluctuationdens}, the limit of integration, $\phi_{\rm
RLP}(Z)$ in Eq. (\ref{compactivity}), changes as discussed in the
phase diagram of \cite{jamming2}, increasing as $\mu$ decreases
and effectively depending on $Z$ as $\phi_{\rm RLP}(Z)$.  Indeed,
$\phi_{\rm RLP}(Z)$ is the theoretical RLP line depicted as a
dashed line in Fig. \ref{phasesim}.  These equations of state
should be compared with an analogous equation of state obtained in
\cite{swinney} for jammed packings equilibrated using a fluidized
bed.  The fluctuations presented in Fig. \ref{fluctuationdens}
monotonically decrease as $\phi$ increases, and do not display a
parabolic shape as depicted in \cite{swinney}.  While the protocol
used herein, the "split" algorithm presented in Section
\ref{simresults}.A \cite{jamming2}, and the protocol for the
experiments of \cite{swinney} differ, we would expect that the
equation of state should be the same.  Elucidation regarding this
difference requires further investigation.

An important note is that this extensive relationship occurs well
before $n$ is large enough such that finite size effects of the
system force the fluctuations to tend to zero.  Further, the
linear relationship extracted from $n$-clusters is different from
that extracted from $n$ randomly chosen Voronoi cells, implying a
correlation between Voronoi cell volumes, revealed using the
clustering technique. Analysis of the fluctuation density reveals
the following formula
\begin{equation}
\begin{split}
\Delta\sigma_n^2 & \equiv \langle ({\cal W}_n - \langle {\cal
W}_n\rangle )^2 \rangle_n = \frac{1}{2}((\sigma_{n+1}^2 -
\sigma_{n}^2) + (\sigma_{n}^2 - \sigma_{n-1}^2)) \\ & = \sigma_1^2
+ \langle {\cal W}_n\Delta {\cal W}_{n+1}\rangle + \langle {\cal
W}_{n-1}\Delta {\cal W}_{n}\rangle
\\ & - \langle {\cal W}_n \rangle \langle \Delta {\cal W}_{n+1}
\rangle - \langle {\cal W}_{n-1} \rangle \langle \Delta {\cal
W}_{n} \rangle \label{corr},
\end{split}
\end{equation}
where $\sigma_1^2$ is the single particle, or microscopic,
fluctuation in Voronoi volume, and $\Delta {\cal W}_{n+1} =
W^{vor}_{n+1} - \langle W^{vor} \rangle$ is the $n+1$ free Voronoi
volume to be added to the cluster of $n$ free Voronoi volumes
being averaged.  If volumes are chosen randomly, as opposed to the
clustering condition, the fluctuation density collapses to the
microscopic fluctuation density, $\sigma_1^2$, as the correlation
tends to $zero$.   Therefore, the fluctuations will scale exactly
with $n$, as indicated in \cite{dauchot}. Further, if the
averaging process is taken to be equal to or larger than the
system size, then $\langle \Delta {\cal W}_{n+1} \rangle = \langle
\Delta {\cal W}_{n} \rangle = 0$, such that Eq. (\ref{corr}) is
rewritten as
\begin{equation}
\Delta\sigma_{n}^2 = \sigma_1^2 + \langle {\cal W}_n\Delta {\cal
W}_{n+1}\rangle + \langle {\cal W}_{n-1}\Delta {\cal W}_{n}\rangle
\label{corr1}.
\end{equation}

Equation (\ref{corr1}) thereby provides an analytical form for the
curves presented in Fig. \ref{fluctuations}, as a function of $n$.
Future work may take into consideration $n$th neighbor
coordination and distance distribution, as in \cite{kumar2}.

{\it Compactivity.---}The compactivity is then obtained via the
integration of Eq. ({\ref{compac1}):

\begin{equation}
X^{-1} = \lambda \int_{\phi(X)}^{\phi_{\rm RLP}(Z)} \frac{d\langle
{\cal W}_n\rangle}{\langle ({\cal W}_n - \langle {\cal W}_n\rangle
)^2 \rangle} \label{compac2},
\end{equation}
where we use that $\phi(X\to \infty) \to \phi_{\rm RLP}$
\cite{jamming2}.  Since Voronoi volumes are additive, $\langle
{\cal W}_n\rangle = \langle {\cal W}\rangle = N V_g/\phi$.  The
fluctuations in Voronoi volume are divided by the number of
grains, $N$, thereby introducing the fluctuation density, shown in
Fig. \ref{fluctuationdens} into Eq. (\ref{compac2}).  Therefore,
the above integration is rewritten as:
\begin{equation}
(X/V_g)^{-1} = \lambda \int_{\phi_{\rm RLP}(Z)}^{\phi(X)}
\frac{d\phi}{\phi^2 \langle ({\cal W}_n - \langle {\cal
W}_n\rangle )^2 \rangle_n} \label{compactivity},
\end{equation}
and we may then utilize the fluctuations as a function of $\phi$,
and integrate along a line of constant $Z(\mu)$.  The assumption
that $\sigma_{n}^2(V)/\langle V \rangle^2 =
\sigma_{n}^2(\phi)/\langle \phi \rangle^2$ is not utilized here,
as done in \cite{swinney}, explaining the different functional
form for the compactivity of Eq. (\ref{compactivity}) from that of
\cite{swinney}.

Following the above presented method, $\phi_{\rm RLP}(Z)$ is
extracted from the phase diagram, and used as a limit of
integration in order to calculate $X(\phi)$ from fluctuations in
Voronoi volume, in Eq. (\ref{compactivity}).  This introduces the
dependence on $Z$ as $\phi(X)$ is obtained by integrating Eq.
(\ref{compactivity}) numerically by applying a fitting function to
the numerical data of Fig. \ref{fluctuationdens}.  The equation of
state, $\phi(X)$ for a given $Z$, is plotted in Fig. \ref{compsim}
for different values of the average coordination number of the
packings, $Z(\mu)$, revealing that as we approach $\phi_{\rm
RCP}\approx 0.645$, $X \to 0$, regardless of the value of $\mu$.
Further, $X \to \infty$ as we approach $\phi_{RLP}$, with the
smallest volume fraction of the RLP appearing for $\mu\to \infty$
and $Z\approx 4$ in the high-compactivity limit, $\phi_{\rm
RLP}\approx 0.55$. The compactivity curve plotted in Fig.
\ref{compsim} is continuous, even though the volume fluctuation
data is the result of a discrete set of simulations as seen in
Fig. \ref{fluctuationdens}. This is due to the fact that we
smoothly interpolate a continuous curve for the volume
fluctuations as a function of $\phi$, resulting in a smooth
integration for the compactivity, and subsequently the entropy
through Eq. (\ref{compactivity}).

\begin{figure}
\centering { \vbox{
 \resizebox{8cm}{!}{\includegraphics{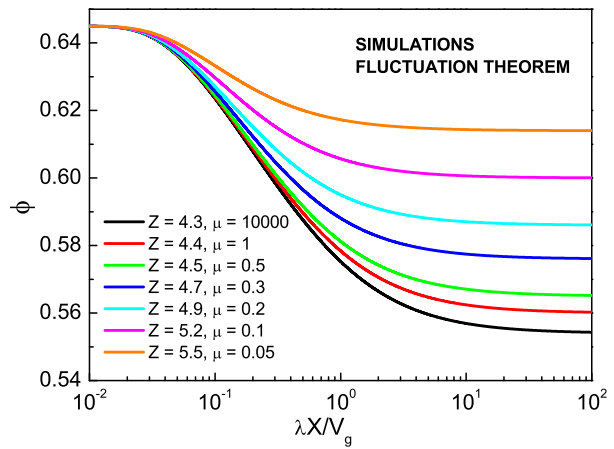}}
} } \caption{$\phi$ versus X from the integration of Voronoi
volume fluctuations.  The smoothness of the curves is due to the
fact that we use fitting functions for the data in Fig.
\ref{fluctuationdens} to perform the integration of Eq.
(\ref{compactivity}) } \label{compsim}
\end{figure}

{\it Entropy.---} The entropy, $S$, and its density, $s=S/N$, are
obtained by integrating

\begin{equation}
X^{-1}= \frac{\partial S}{\partial V}.
\end{equation}

By virtue of having a fixed total volume, $V$, for any particular
system defined by $\phi$, we can substitute $V = N V_g/\phi$ such
that:

\begin{equation}
(X/V_g)^{-1}= -\phi^2 \frac{\partial s}{\partial\phi}
\end{equation}

Using the concept that $\phi_{\rm RCP}$ is a fixed value in the
phase diagram for all values of $Z(\mu)$, we integrate between the
limits of $\phi_{\rm RCP}$ and the desired $\phi$.

\begin{equation}
s(\phi) - s(\phi_{\rm RCP}) = \lambda \int^{\phi_{\rm RCP}}_{\phi}
\frac{d\phi}{(\lambda X/V_g)\phi^2}. \label{entropy}
\end{equation}

Equations (\ref{compactivity}) and (\ref{entropy}) can be combined
to provide an equation for $s$ as a function of the Voronoi volume
fluctuations.

\begin{equation}
s(\phi) =  \lambda \int^{\phi_{\rm RCP}}_{\phi}
\frac{d\phi'}{\phi'^2} \int_{\phi_{\rm RLP}(Z)}^{\phi'}
\frac{d\phi''}{ \phi''^2\langle ({\cal W}_n - \langle {\cal
W}_n\rangle )^2 \rangle} \\ +  s(\phi_{\rm RCP})
\label{singleentropy}.
\end{equation}

Therefore, we can calculate $s(\phi)/\lambda$, based on the
fluctuations of a packing of jammed grains for a fixed $Z$,
following the horizontal lines in Fig. \ref{phasesim}. Integration
of the compactivity curve achieved via simulation provides the
entropy, up to a constant value at $\phi_{\rm RCP}$, as defined by
Eq. (\ref{entropy}).  The entropy of the packings from simulations
in Fig. \ref{phasesim} is plotted in Fig. \ref{entsim} as a
function of $\phi$ for different values of $Z$.  The shape of the
entropy curve is in qualitative agreement with that calculated by
Aste and collaborators using X-ray tomography techniques to
determine the position of particles inside large scale packings,
as shown in \cite{aste3}.

\begin{figure}
\centering { \vbox{
 \resizebox{8cm}{!}{\includegraphics{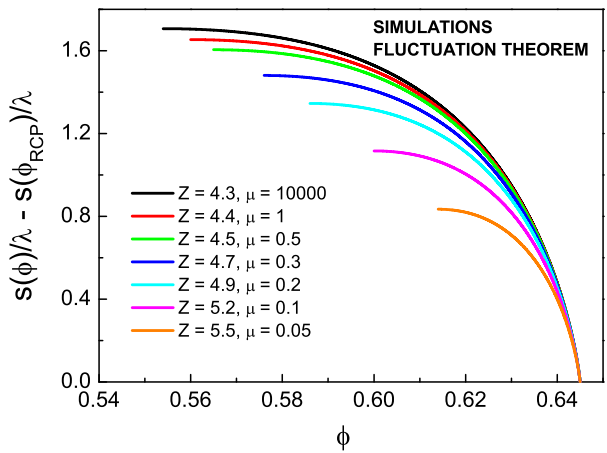}}
} } \caption{Entropy density versus $\phi$ from the integration of
X.} \label{entsim}
\end{figure}

\subsection{Entropy from Information Theory}

Analysis of the entropy from fluctuations in Voronoi volume
clusters provides a value for entropy density up to a constant of
integration $s(\phi_{\rm RCP})$, as shown in Eq.
(\ref{singleentropy}). To obtain the entropy of RCP we use an
independent method based on information theory as developed in
\cite{shannon,vink}, related to the thermodynamic entropy in
\cite{kumar1}, and applied to emulsion systems in \cite{brujic}
which does not require a constant of integration. This method
provides a second estimate of the entropy for all volume fractions
to be compared with Fluctuation theory.

We use the Voronoi cell and Delaunay triangulation for each
particle to define a Voronoi network by considering contacts when
a Voronoi side is shared between two particles, and hence are
Delaunay contacts, as shown in Fig \ref{polygon}.  In order to
facilitate periodic boundary conditions, we surround the finite
box enclosing all Voronoi cells with $26$ virtual boxes.  These
boxes enclose virtual particles, translated in all possible
combinations from the real box.  QHull calculates all Delaunay
contacts, and only those pairs of contacts which contain at least
one real particle are considered, while pairs of virtual contacts
are discarded, thus ensuring complete periodic boundary
conditions.

A graph is constructed as a cluster of $n$ particles that are
Delaunay contacts, and by means of graph automorphism \cite{mckay}
can be transformed into a standard form or "class" $i$ of
topologically equivalent graphs that are considered a state with a
probability of occurrence $p_i$. Reference \cite{mckay} provides
the code to deal with isomorphic graphs, which is essential for
counting different graph classes.  The topological equivalence
accounts for graphs with varied Voronoi cell sizes and locations
while retaining identical lists of consecutive subsets of Delaunay
contacts.  In practice, we determine $p_i$ by extracting a large
number $m$ of clusters of size $n$ from the system and count the
number of times, $f_i$, a cluster $i$ is observed, such that:

\begin{equation}
p_i=f_i/m. \label{prob}
\end{equation}

A simple example of graph classes can be seen when $n = 3$. Figure
\ref{shannon3} shows exactly $4$ possible graphs that can be
achieved, consisting of $0, 1, 2$ and $3$ connections between $3$
Delaunay contacts.  Each graph falls into a class of $i = 0
\rightarrow 3$. The random process of selecting a graph with $n=3$
is done $m$ times, continuously calculating the probability of
$f_i$, as defined in Eq. (\ref{prob}).

\begin{figure}
\centering { \vbox{
 \resizebox{6cm}{!}{\includegraphics{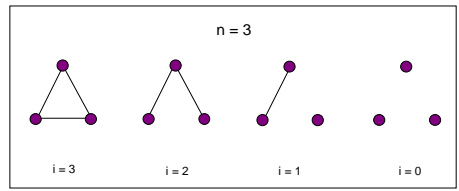}}
} } \caption{Classes of graphs for n=3.  There are only 4 possible
graphs, 0 contacts, 1 contact, 2 contacts, or 3 contacts.}
\label{shannon3}
\end{figure}

The Shannon entropy of a clusters of size $n$ is thereby defined
as:

\begin{equation}
H(n)=-\lambda \sum p_i \ln p_i. \label{shannon0}
\end{equation}

Equation (\ref{shannon0}) reduces to the thermodynamic entropy if
one replaces the probability, $p_i$, with the Boltzmann factor.
The name Shannon entropy does not imply that Eq. (\ref{shannon0})
is a new kind of entropy.  It merely implies that we will
calculate the entropy of the packing using information theory
methods, which is routinely applied to sequences.

Ideally, $m$ is large enough such that the values of $p_i$
converge to a fixed value for each value of $n$, such that the
Shannon entropy converges as well. Computationally, for large $n$,
it is not possible to reach this convergence within a reasonable
time, and certain approximations can be applied as outlined in
\cite{rombouts} to decrease the number of iterations necessary,
summarized briefly here.

As one iteratively selects clusters, increasing $m$ by one each
time, new classes of graph are obtained, such that we have a total
number of different graphs $k$, where $k \leq m$.  While Fig.
\ref{shannon3} depicts the relatively simple case of $n=3$, higher
values of $n$ have a significantly larger number of graph classes.
Therefore, Eq. (\ref{prob}) is only an approximation of the true
probability of observing a cluster $i$ and can be rewritten as
\begin{equation}
p_i \approx f_i/m, \label{prob1}
\end{equation}
where the equation becomes an equality as $m\rightarrow\infty$. We
must therefore replace Eq. (\ref{shannon0}) by

\begin{equation}
H^{*}(n)=-\lambda \sum \frac{f_i}{m} \ln
\frac{f_i}{m}\label{shannon1}.
\end{equation}

Graphs with $f_i/m$ smaller than $1/m$ will likely be observed
only once, if at all. This finite-size effect grows with
increasing $n$, as graphs become very complex. The quantity
$H_{1}(n)$ is defined as the contribution to $H^{*}(n)$ of the
topologies observed once. Measurements where $H_{1}(n)$ exceeds a
threshold percentage of $H^{*}(n)$ are not considered to be valid
measurements.

In an effort to improve convergence, and thereby decrease
simulation time for the calculation of the configurational
entropy, a finite sample correction is applied.  The details of
this correction are presented herein, as well as in
\cite{rombouts}.

Referring now to $H^*$ from above, the probability $P_i(f_i)$ that
a certain state $i$ will be observed exactly $f_i$ times is given
by the binomial distribution

\begin{equation}
P_i(f_i) = {m \choose f_i} p_i^{f_i} (1-p_i)^{m-f_i}.
\end{equation}

Define the probability of a certain event to be $p_i$, the number
of observed events to be $m$ and the number of uniquely observed
events to be $k$. Furthermore let $F(x)$ be a function that can be
Taylor expanded around $p_i$:

\begin{equation}
F(x) = F(p_i) + F'(p_i)(x-p_i) + \frac{1}{2}F''(p_i)(x-p_i)^2 +...
\end{equation}

The binomial distribution is concentrated around the average
$\langle f_i/m \rangle = p_i$, and we obtain the following useful
approximation using the definition of the variance in a binomial
distribution:

\begin{equation}
\begin{split}
\langle F(\frac {f_i}{m}) \rangle & = F(p_i) + F'(p_i)\langle
\frac {f_i}{m}-p_i \rangle + \frac{1}{2}F''(p_i) \langle (\frac
{f_i}{m}-p_i)^2 \rangle +... \\ & = F(p_i) + \frac{1}{2}F''(p_i)
\frac {p_i(1-p_i)}{m} + ...
\end{split}
\end{equation}

Let $F(x) = -x \log x$, an obvious choice considering the form of
the Shannon entropy.  Then, $F''(x)$ = $-1/x$ and

\begin{equation}
\langle F (\frac{f_i}{m}) \rangle = -p_i \log p_i -
\frac{1-p_i}{2m},
\end{equation}

Therefore,

\begin{equation}
\begin{split}
\frac{\langle H^* \rangle}{\lambda} & = \sum_i \langle
F(\frac{f_i}{m})
\rangle =-\sum_i p_i \log p_i - \frac {\sum_i (1-p_i)}{2m} +... \\
& = \frac{H}{\lambda} - \frac{k-1}{2m} +...,
\end{split}
\end{equation}

The Shannon entropy is then the calculated entropy from the
average of $H^*$ obtained from multiple simulations, plus the
binomial correction term $(k-1)/2m$, since the sum over all $p_i$
is unity, a term clearly tending to zero as $m$ tends to infinity.
This approximation works well under two conditions. First, $mp
\geq 1$, in order to allow the Taylor expansion to converge.
Second the contribution of the probabilities below $1/m$ to the
Shannon entropy must be a practically insignificant term. If this
is not the case, additional binomial correction terms must be
used. However, these terms will not be independent of $p_i$,
therefore making the calculation significantly more complicated.
It is of interest in the present work to ensure that both
conditions necessary for the use of only the first term of the
binomial correction are applicable.

Assuming the above described binomial distribution of the
probability of having a cluster of class $i$ observed exactly
$f_i$ times, the first order finite sampling correction to $H^*$
therefore results in

\begin{equation}
\frac{H(n)}{\lambda} = \frac{H^*(n)}{\lambda} + \frac{k-1}{2m}.
\end{equation}

The entropy must be further corrected to account for the Shannon
entropy as measured for a crystal structure by using the methods
presented herein.  The entropy of a crystal structure where
$N\rightarrow\infty$ should be zero, and as such an FCC packing
should have zero entropy.  However, when applying the graph
counting method explained above to a finite system, a crystal
structure will present a non-zero entropy.  It is important to
subtract the entropy of a finite crystal structure from the
Shannon calculations.  In studies of network forming materials
outlined by \cite{vink}, an empirical correction term of $g(n) =
(d-1) \log(n)$ is subtracted from each value of $H(n)$, where $d$
is the dimensionality of the network. The functional form of
$g(n)$ is obtained by applying the above described Shannon entropy
calculation directly to an FCC packing, and its results are shown
in Fig. \ref{crystal}. During the process of randomly selecting
$m$ points in the network, some points will inevitably fall inside
of the grain boundaries. Points approaching the center of a grain
will have a different set of $n$ nearest grain centers then points
chosen outside of the grain boundary.  This discrepancy is
accounted for via $g(n)$.  In the present study, this term is
augmented by using the exact values obtained for $g(n)$, and not
the empirical form, as shown in Fig \ref{crystal}. It should be
noted that the empirical form for $g(n)$ comes very close to the
exact values, differing only by an approximately exact constant.
For calculations of entropy density, differences between
successive values of $H(n)$ result in the cancellation of this
constant, exemplified below.

\begin{figure}
\centering { \vbox{
 \resizebox{8cm}{!}{\includegraphics{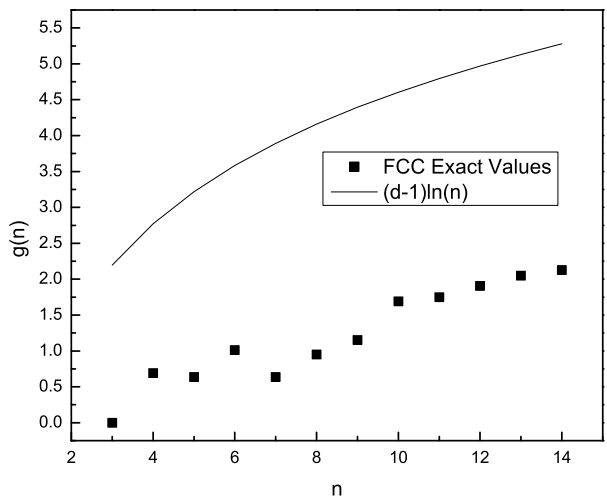}}
} } \caption{Entropy of crystal (FCC) packing as calculated using
graph theoretical methods.  While the entropy is approximately
equal to $(d-1)\ln(n)$, minus a meaningless constant, exact values
from simulations are used herein.} \label{crystal}
\end{figure}

We redefine the Shannon entropy as follows:

\begin{equation}
\frac{H'(n)}{\lambda} = \frac{H(n)}{\lambda} - g(n),
\end{equation}
and the entropy density is
\begin{equation}
s=\lim_{n \to \infty} [H'(n+1)-H'(n)]
\end{equation}
Figure \ref{shannonent} shows the calculation of $H'(n)$ versus
$n$ for a typical packing with $\mu = 10000$ and $\phi = 0.64$. We
show that $s(n)$ converges so rapidly, as shown in Fig.
\ref{shannonent} such that even moderate values of $n$ ($n \geq
8$) are enough to obtain a sufficient approximation of $s$
\cite{vink}. The calculation is repeated for all the packings and
the Shannon entropy density is plotted in Fig.
\ref{shannonentdens} versus $\phi$ for different $Z(\mu)$.

When examining values of the Shannon entropy for values of $\phi <
\phi_{RCP}$, Fig. \ref{shannonentdens} displays an increase in the
entropy as volume fraction decreases, similar to that of the
entropy as calculated via Voronoi volume fluctuation density in
Fig \ref{entsim}. However, this increase does not appear dependent
on the mechanical coordination number of the packings.  Further,
the Shannon entropy does not increase by the same magnitude as
observed in the entropy from fluctuations.  For instance, from the
fluctuation theorem calculations of Fig. \ref{entsim} we find
$s_{\rm RLP} - s_{\rm RCP} = 1.7\lambda$ while from the Shannon
entropy of Fig \ref{shannonentdens} we find $s_{\rm RLP} - s_{\rm
RCP} = 0.35\lambda$. As suggested in \cite{rombouts}, this
discrepancy may be due to the additional entropy arising from
freedom to move grains within the packing without disrupting the
Delaunay network, and hence not affecting the probabilities in Eq.
(\ref{prob1}). Analysis of such volume contributions to the
Shannon entropy requires a Monte Carlo simulation, in which the
available phase space volume that the packings can explore for a
fixed graph is probed \cite{rombouts}, with the additional
constraint that the packing must maintain the Delaunay contact
network under all possible rearrangements. This calculation is
considered in \cite{rombouts} and will be the topic of future
study.

Figure \ref{shannonentdens} shows that as we approach $\phi_{\rm
RCP}$ for all values of $Z$ using information theory, $s/\lambda
\simeq 1.1$. We therefore define $s(\phi_{RCP})/\lambda = 1.1$,
the value of the entropy as calculated via graph theoretical
methods.

Thus, the Shannon entropy density provides an estimation of the
entropy for the RCP state, $s(\phi_{\rm RCP})\approx 1.1 \lambda$,
serving as the constant of integration for the entropy density as
realized by Fluctuations Theorem.  Under this approximation, we
can shift the Fluctuation Theorem entropy of Fig. \ref{entsim}
vertically by $s_{\rm RCP} = 1.1\lambda$ as calculated via Shannon
entropy methods.  Figure \ref{entshift1} shows the entropy shifted
by this constant value, as calculated via simulation.  It is
important to emphasize that $s(\phi_{\rm RCP})$ is approximate,
due to neglecting fluctuations of the fixed Delaunay network as
explained above.  Nevertheless, we remark that the obtained value
of $s_{\rm RCP}$ is compatible with other estimates which found
the entropy of the order $\lambda$ (see for instance the
calculation of the analogous complexity, $\Sigma$, by Zamponi and
Parisi who found $\Sigma / \lambda \sim 1$ \cite{parisi}).

\begin{figure}
\centering { \vbox{
 \resizebox{8cm}{!}{\includegraphics{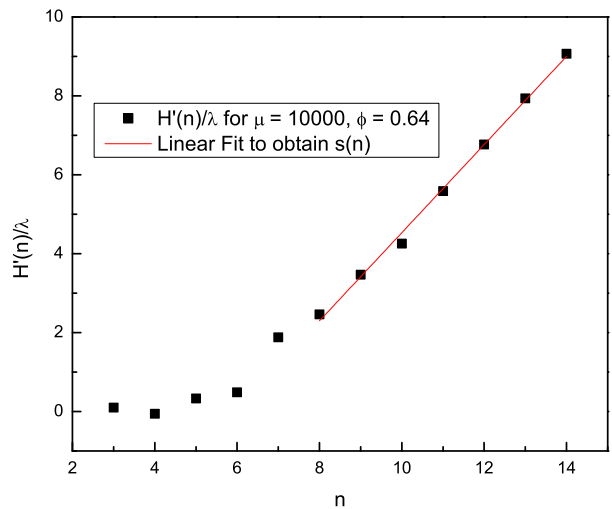}}
} } \caption{Shannon entropy function of n, with $\mu = 10000$ and
$\phi =0.64$.  The red line displays a linear fit between $n = 8$
and $n = 14$, from which the entropy density is extracted.  This
process is repeated for all packings used herein.}
\label{shannonent}
\end{figure}

\begin{figure}
\centering { \vbox{
 \resizebox{8cm}{!}{\includegraphics{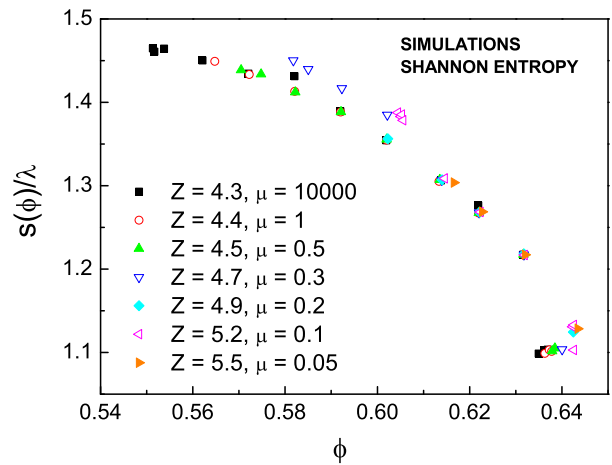}}
} } \caption{Shannon entropy density, $s(\phi)$ for all packings
used herein.  The minimum value of the entropy density is achieved
at RCP, and is used as a constant of integration for the entropy
obtained from Fluctuation theorem.  } \label{shannonentdens}
\end{figure}

\begin{figure}
\centering { \vbox{
 \resizebox{8cm}{!}{\includegraphics{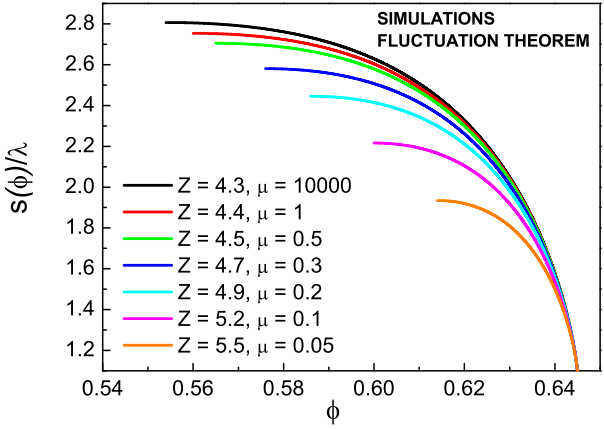}}
} } \caption{Entropy at RCP achieves a value of 1.1 as calculated
by the Shannon Entropy at RCP and serves as a constant of
integration for the entropy from Fluctuation theorem in Fig.
(\ref{entsim}) producing the entropy shown in this figure.  }
\label{entshift1}
\end{figure}

\section{Theoretical Model}\label{tmodel}

In this section we develop a theoretical framework to understand
the numerical results in light of \cite{jamming2}.

\subsection{Statistical mechanics of frictional hard spheres}
Experiments of shaken grains, fluidized beds and oscillatory
compression of grains
\cite{chicago2,swinney,dauchot,maksekurchan,bruijicwang} indicate
that granular materials show reversible behavior, and the analogue
of the conserved energy, $E$, in thermal systems is the volume
$V=N V_g/\phi$, for a system with $N$ grains of volume $V_g$ at
position $\vec{r}_i$.  Thus, the number of configurations,
$\Omega$, and the entropy in the micro-canonical ensemble of
jammed hard spheres is defined as \cite{sirsam}:
\begin{equation}
\Omega(V) = e^{S(V)/\lambda} = \int \delta \Bigl(V-{\cal
W}(\vec{r}_i)\Bigr) \,\,\,\, \Theta_{\rm jam}(\vec{r}_i)  \,\,
{\cal D}\vec{r}_i. \label{omega}
\end{equation}
Analogous to the temperature in equilibrium system $\partial
E/\partial S = T$, the ``temperature'' in granular matter is the
compactivity, $X=\partial V/\partial S$.  Here $\Theta_{\rm
jam}(\vec{r}_i)$ is a constraint function restricting the integral
to the ensemble of jammed states, ${\cal W}(\vec{r}_i)$ is the
volume function associated with each particle taking the role of
the Hamiltonian in thermal systems which is defined in terms of
the Voronoi volume in Eq. (\ref{vor1}). The crux of the matter is
then to properly define $\Theta_{\rm jam}$ and $\cal W$ to
calculate the entropy and volume in the ensemble of jammed matter.

{\it Volume ensemble.---} A minimum requirement of $\Theta_{\rm
jam}(\vec{r}_i)$ is to ensure touching grains, and obedience to
Newton's force and torque laws. As in the numerical simulations,
the volume function, ${\cal W}(\vec r_i)$, is taken as the volume
of the Voronoi cell associated with each particle at position
$\vec r_i$, for which the analytical form has been obtained in Eq.
(\ref{vor1}) \cite{jamming2}. The entropy in the V-ensemble of
frictional hard spheres is:
\begin{equation}
\begin{split}
  e^{S(V)/\lambda} &= \int
  \delta \Bigl(V- {\cal W}(\vec{r}_i) \Bigr)  \times \\
  \prod_i \Bigl\{ \delta \Bigl(\sum_{j \neq i} \vec{f}_{ij} \Bigr) &
  \delta\Bigl(\sum_{j \neq i} \vec{f}_{ij} \times \vec{r}_{ij}\Bigr)
  \,\, \delta(\vec
  f_{ij} - \vec f_{ji}) \times \\
  \prod_{j \neq i} \Bigl[
  \Theta(\mu f_{ij}^{\ N} - f_{ij}^{\ T})  &
\delta \left([(\vec
    r_{ij})^2-1](\vec f_{ij})^2\right)
 {\cal D} \vec f_{ij}
  \Bigr] {\cal D} \vec r_i \Bigr\},
\end{split}
\label{exact}
\end{equation}
where $\vec r_{ij} \equiv \vec r_i - \vec r_j$, the normal
inter-particle force is $f_{ij}^{\ N} \equiv |\vec f_{ij}\cdot
\hat r_{ij}|$, the tangential force: $f_{ij}^{\ T} \equiv |\vec
f_{ij} - (\vec f_{ij}\cdot \hat r_{ij})\hat r_{ij}|$. All
quantities are assumed properly a-dimensional for simplicity of
notation. The terms inside the brackets $\{\cdot\}$ correspond to
the jamming constraint function $\Theta_{\rm jam}$ in Eq.
(\ref{omega}), and therefore define the ensemble of jammed states.
The first three $\delta-$functions inside the big brackets impose
Newton's second and third law. The Heaviside $\Theta-$function
imposes the Coulomb condition and the last $\delta-$function the
touching grain condition for hard spheres, assuming identical
grains of unit radius.  Integration is over all forces and
positions which are assumed to be equally probable as in the flat
average assumption in the micro-canonical ensemble.

An extra term should be added as $
\delta\left(\frac{1}{2V}\sum_{i\neq j}(
\vec{f}_{ij}\otimes\vec{r}_{ij})-\bar{\bar{\sigma}}\right)$ where,
for the isotropic case, the stress tensor is $\sigma_{\alpha\beta}
= p\delta_{\alpha\beta}$, with $p$ the pressure. Since we are
treating hard spheres, the pressure $p$ just controls the mean
value of forces and does not contribute to the statistics. Thus,
this term is not needed, which means that the angoricity,
$A=\partial p/\partial S$, is irrelevant for hard spheres.  At the
isostatic limit $p \rightarrow 0$, and further assuming a
Boltzmann distribution of pressure
\cite{blum1,blumed,chak,serb,vanhecke}, similar to that of
mesoscopic volumes \cite{jamming2}, such a term would tend towards
unity in a partition function for jammed matter. However, the
angoricity should be considered in the case of deformable grains,
a system of future studies.

Clearly, Eq. (\ref{exact}) is almost intractable from an
analytical point of view.  However, under the quasi-particle
approximation of \cite{jamming2} we can define the configurational
entropy at the mesoscopic level using a corollary of the
force-balance ensemble: the isostatic conjecture and a
coarse-grain volume function in terms of the coordination number
as we show in Section \ref{tmodel}.B.

It is of interest to determine if the ensemble defined by Eq.
(\ref{exact}) satisfy the definition of jamming given by Torquato
\cite{salvatore1}. A definition of jammed configurations based on
force/torque balance is a necessary condition for mechanical
equilibrium of a packing with friction and frictionless grains. A
force-balanced packing is defined as the existence of a set of
forces $\{\vec{f}_{ij}, \forall\ \mathrm{balls}\ i$ in contact
with ball $\alpha\}$ such that the sum of the forces/torques for
each particle is zero,
\begin{equation}
\sum_j \vec{f}_{ij} = 0, \,\,\,\,\,\,\,\, \sum_j
\vec{f}_{ij}\wedge\hat n_{ij} = 0, \label{r1}
\end{equation}
with the non trivial contact forces $\sum_{ij}|\vec{f}_{ij}| = 1,
\vec{f}_{ij}\cdot \hat{n}_{ij}\geq 0$ where $\hat{n}_{ij}$ is the
unit vector of the contact $j$. Other necessary mechanical
constraints should be added too, e.g.
$\vec{f}_{ij}\wedge\hat{n}_{ij} = 0$ for frictionless packings.

While the above conditions are necessary for jamming they are not
sufficient. To  define a more restricted force-balance condition
we  consider that  the contacts around a ball $i$ are \emph{not
degenerated} if $\exists$ a neighbor $\alpha$ such that
\begin{equation}
\vec{f}_{i\alpha}\cdot\hat{k} \neq 0, \,\,\,\, \,\,\,\,\,\,\,
\forall \hat{k} \neq 0. \label{r2}
\end{equation}
This condition assures that at least one contact force is
off-plane. A packing is {\it restrict force-balanced} if it is
force-balanced and the contacts around all the balls are not
degenerated.

A question is raised whether the restrict force-balance condition
for the frictionless case fits the geometrical definition of
Torquato \cite{salvatore1}. It can be proved that packings with
(\ref{r1})  and (\ref{r2}) are at least locally jammed. To see
this, let us formalized the locally jammed condition as follows:
for any ball $i$, there is at least one contact
$\hat{n}_{i\alpha}$ such that $\hat{n}_{i\alpha}\cdot\hat{k}
> 0$, $\forall \hat{k} \neq 0$. Since the contact force
$\vec{f}_{i\alpha}$ is parallel with $\hat{n}_{i\alpha}$ for
frictionless packings, the condition is equal to
$\vec{f}_{i\alpha}\cdot\hat{k} < 0$. From the non-degenerated
condition, there must be a contact satisfying
$\vec{f}_{i\alpha}\cdot\hat{k} \neq 0$. Assuming
$\vec{f}_{i\alpha}\cdot\hat{k}
> 0$ (otherwise we prove the result directly), there must exist
another contact $\alpha'$ that $\vec{f}_{i\alpha'}\cdot\hat{k} <
0$ because of the force-balance condition, then $\sum_{j}
\vec{f}_{i\alpha}\cdot\hat{k} = 0$.

Thus we prove a sufficient condition for local jammed
configuration from the force balance point of view.  Accordingly,
the non-degenerated condition is necessary for the proof above.
Since it not necessary it could also satisfy more stringent
geometrical definitions, e.g. collective jammed, or even strict
jammed, but this  has to be figured out in further investigation.
Although the restrict force-balance condition is not always
satisfied in our simulations, the bias is very small. In
frictionless packings it implies that the coordination number must
be larger than $d$; a small fraction of particles with $Z<d$ are
found in simulated packings as well as experimental ones. By
removing the degenerated balls recurrently (for instance it is a
common practice to remove floaters), we end up with a packing
satisfying the constraint of non-degeneracy and therefore being
locally jammed. Thus we expect that many experimental packings
satisfy the restrict force-balance condition. The force-balance
condition can be extended to the frictional cases, by simply
adding the constraint between normal and tangential forces.
Finally, the ensemble in Eq. (\ref{exact}) is assumed to satisfy
the restrict force-balance condition.  Thus, it is understood that
the condition of Eq. (\ref{r2}) is implicit in the ensemble
average of Eq. (\ref{exact}).

Assuming that the conditions specified in Eq. (\ref{exact}) are
met in the numerical packings, the simulation results can be
interpreted as the ensemble average Eq. (\ref{exact}). However,
there is a further important distinction between Eq. (\ref{exact})
and the numerical calculations.  Equation (\ref{exact}) assures
that all configurations at a given volume have the same
probability. This is the flat average assumption in the
micro-canonical ensemble that allows for the development of
statistical mechanics.  Without this assumption statistical
calculations are impossible to perform (see however
\cite{dauchot1} for a thorough discussion). There is no rigorous
proof that the flat average assumption is correct in equilibrium
statistical mechanics. Still, its validity is widely accepted. For
granular matter, use of the flat average is much more
controversial.  Earlier simulations \cite{maksekurchan} indicate
some evidence for ergodicity.  The ergodic hypothesis implies not
only the equal probability of states but also that for
sufficiently long times the phase trajectory of a closed system
passes arbitrarily close to a manifold defined by a constant
volume (or energy) \cite{landau-stat-mech}. Experiments indicate
reversible behavior, supporting that the flat average can be
applied to granular matter, although this assertion is certainly
not true in general.

We notice that Eq. (\ref{exact}) is difficult to solve. Analytical
progress can be done by considering a coarse-graining of the
Voronoi volume function and working with quasi-particle theory
developed in \cite{jamming2} to obtain a configurational entropy
at the mesoscopic level.

\subsection{Volume Function} The mesoscopic theory presented in
\cite{jamming2} and \cite{jamming1} coarse grains the Voronoi
volumes of a jammed granular packing over a mesoscopic length
scale and calculate an average volume function. The coarsening
reduces the degrees of freedom to one variable, the geometrical
coordination number of each grain, $z$, that allows for an
analytical solution of the partition function.  We find a
mesoscopic free volume function \cite{jamming2}:
\begin{equation}
 w(z) \equiv \frac{\langle {\cal W}_i^{vor} \rangle - V_g}{V_g} = \frac{2\sqrt{3}}{z},
\label{hamil}
\end{equation}

Here, we note that ${\cal W}_i^{vor}$ has been rigorously defined
in Eq. (\ref{vor1}), validated its equivalence to the Voronoi
volume, and its use in comparison between simulation results and a
statistical mechanics formulation utilizing the Voronoi cell as
the microscopic volume associated with each grain.

If the system is fully random we can extend the assumption of
uniformity from the mesoscopic scale to the macroscopic scale,
such that we arrive to an equation of state relating $\phi^{-1}= w
+ 1$ with $z$ as:

\begin{equation}
\phi=\frac{z}{z+ 2\sqrt{3}}. \label{phi4}
\end{equation}

\subsection{Partition Function}

Below, we briefly discuss the results of \cite{jamming2} regarding
the phase diagram.

We assume that the sum over each quasiparticle with volume $w(z)$
is the total volume \cite{jamming1}, then the canonical partition
function for a single particle can be written as:

\begin{equation}
\label{partw} {\cal Z}_{\rm iso} = \int g(z)
e^{-\frac{w(z)}{\lambda X}} dz
\end{equation}

Equation (\ref{partw}) is the single particle partition function,
such that the full partition function for $N$ particles is ${\cal
Z}_{\rm iso}^N/N!$.

The density of states, $g(z)$ is assumed to be analogous to the
result when the discreteness of phase space imposed by the
Heisenberg uncertainty principle in quantum mechanics.  We assume
the density of states to be only a function of the geometrical
coordination number, as the volume function from Eq. (\ref{hamil})
reduces the degrees of freedom for jammed granular matter to $z$.
The density of states is thereby conjectured to take the form
\cite{jamming2}:
\begin{equation}
g(z) = (h_z)^{z-2d} \label{gz},
\end{equation}
where $h_z$ is the analogue of the Planck constant (see
\cite{jamming2} for more details).

The most populated state is the highest volume at $z=4$ while the
least populated state is the ground state at $z=6$.  This assumed
form of Eq. (\ref{gz}) is an approximation, and will be further
addressed below. Since the term $1/(h_z)^{2d}$ is a constant, it
will not influence the average of the observables in the partition
function, although it changes the value of the entropy by a
constant independent of $\phi$.

Conditions of isostaticity provide the lower bounds of the
geometrical coordination number as $Z \leq z$, while considering
disordered states and hard sphere conditions imposes $z \leq 6
\leq 2d$. which induce bounds upon the limits of integration in
the partition function, and account for the jamming restriction,
$\Theta_{\rm jam}$. Here, $Z$ is the mechanical coordination
number related to the force balance (that is, counting the
contacts with non-zero forces) while $z$ is the geometrical
coordination number related to the geometry of the contact
network.  $Z$ ranges between $Z = d+1$, for infinite friction
grains, and $Z = 2d$, for frictionless grains as given by the
isostatic conjecture discussed by Alexander \cite{alexander} and
in many subsequent papers.   Further detail on this notion is
available in \cite{jamming2}.

Substituting Eq. (\ref{hamil}) and Eq. (\ref{gz}) and the
isostatic condition into Eq. (\ref{partw}), we find the isostatic
partition function:

\begin{equation} \label{pf1}
{\cal Z}_{\rm iso}(X,Z) = \int_{Z}^6 (h_z)^{z-2d}
\exp\left(-\frac{2 \sqrt{3}}{z X}\right ) dz.
\end{equation}

Obtaining the phase diagram is then a matter of calculating the
average volume fraction, $\phi(X,Z)$ by solving the partition
function for different values of $X$ and $Z$.

\begin{eqnarray}
\begin{split}
&\phi(X,Z)=&\\
&\frac{1}{{\cal Z}_{\rm iso}(X,Z)} \int_{Z}^6
\frac{z}{z+2\sqrt{3}} \exp\left(-\frac{2 \sqrt{3}}{z X}+z \ln
h_z\right)dz.&
\end{split}
\label{phi1}
\end{eqnarray}

The boundaries of the phase diagram are plotted in Fig.
\ref{phasesim} along with the packings obtained via simulation.
We note that $Z_{\rm RCP}$ from simulations appears slightly
higher than $Z = 6$, falling closer to $Z_{\rm RCP} = 6.2$ as
shown in Fig. \ref{phasesim}.  We will use this value as $Z_{max}
= 6.2$ when considering it in the partition function as an upper
bound for the jamming condition in an effort to accurately analyze
the simulation results and characterize the entropy. Further,
since the variation in $Z$ is small as we examine lines of
constant $\mu$, the average of $Z({\mu})$ can be used as the lower
bound of the limit in the partition function.

\subsection{Compactivity}

The results presented in Fig. \ref{compsim} are compared to the
theoretical model presented above.  The theoretical calculation
for $\phi(X)$ is achieved exactly as described for $\phi(Z)$ in
Eq. (\ref{phi1}), and is presented in Fig. \ref{comptheory}.

\begin{figure}
\centering { \vbox{
 \resizebox{8cm}{!}{\includegraphics{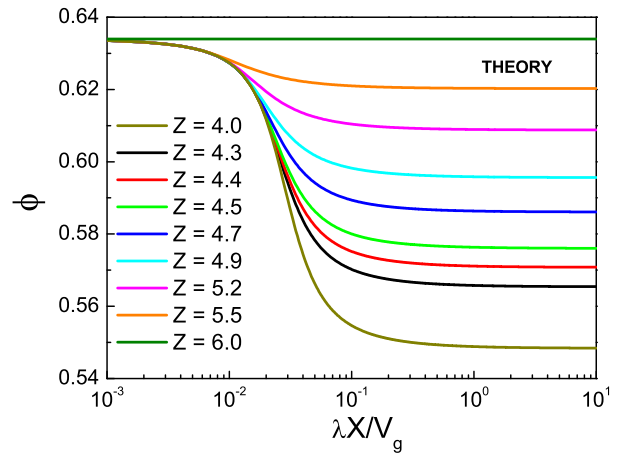}}
} } \caption{Prediction of the mesoscopic theory for $\phi(X)$.}
\label{comptheory}
\end{figure}

In the limit of vanishing compactivity ($X\to 0$) for the
theoretical model, only the minimum volume or ground state at
$z=6$ contributes to the partition function. Then we obtain the
RCP state, $\phi_{\rm RCP} = \phi(X=0,Z) = \frac{6}{6+2\sqrt{3}}
\approx 0.634$, for all values of $Z$. In the limit of infinite
compactivity ($X \to \infty$), the Boltzmann factor $e^{\frac{-2
\sqrt{3}}{z X}} \to 1$, and the average in (\ref{phi1}) is taken
over all the states with equal probability. Assuming $h_z\ll 1$,
the leading contribution to the average value is from the highest
volume at $z=Z$ and therefore we obtain
\begin{equation}
\phi_{\rm RLP}(Z) \approx \frac{Z}{Z+2\sqrt{3}}.
\label{phirlpapprox}
\end{equation}

The dotted line in Fig. \ref{phasesim} is a plot of the equation
of state presented in Eq. ({\ref{phirlpapprox}).  It is an
important result that the values of $\phi_{\rm RLP}(Z)$ well fit
$\phi = \frac{Z}{Z+2\sqrt{3}}$.  On the other hand, there are
states to the right of the RCP line in Fig. \ref{phasesim}.  These
states are a manifestation of the microscopic fluctuations, and
not taken into account by the present mesoscopic theory.

The general shape of $\phi(X)$, achieved via simulation well
matches that as predicted by the mesoscopic theory, if not in
magnitude, when using an exponential form for $g(z)$.  The
compactivity achieves maximal value at the minimal available
volume fraction, as observed both when examining the fluctuations
in Voronoi volume and within the confines of the mesoscopic
theory.  Packings near RLP, being the least dense, have the
greatest room to be further compacted, or increase density, and
therefore approach infinite compactivity. Packings near RCP, being
the most dense, have either minimal or zero room for to be further
compacted, and therefore cannot increase density and tend towards
zero compactivity.  This helps to establish the concept of
compactivity as a static "effective temperature", acting as a
state variable that may link the results of packings preparation
and specific packing protocols to the statistical mechanics
formulation.  Indeed, these results for $\phi(X)$ qualitatively
resemble the compaction curves obtained in the experiments of
\cite{chicago2}.  We also note that the results of Fig.
\ref{comptheory} are qualitatively similar to a simple mean-field
two state model predicted by Edwards, where RCP and RLP are
obtained in the limits of $X\rightarrow0$ and
$X\rightarrow\infty$, respectively \cite{sirsam}.

\subsection{Entropy}

Comparison to the theoretical model proceeds by defining the
equation of state for the entropy density.  The entropy density
is obtained as:

\begin{equation}
s_{\rm meso}(X,Z) = \langle w \rangle /X + \lambda \ln {\cal
Z}_{\rm iso}(X,Z) \label{eqnofstate}
\end{equation}

This equation is obtained in analogy with equilibrium statistical
mechanics and it is analogous to the definition of free energy:
$F=E - T S$ where $F=- k_B T \ln {\cal Z}$ is the free energy. We
replace $k_B T\rightarrow \lambda X$, $E\rightarrow
\left<w\right>$. Therefore, $F = E - TS$ or $S = (E-F)/T = E/T +
\ln {\cal Z} $ is now $s(X,Z) = \left<w\right>/X + \lambda \ln
{\cal Z}_{\rm iso}(X,Z)$. The partition function is evaluated by a
numerical integration of Eq. (\ref{pf1}) for a fixed $Z$ and as a
function of $X$.  A numerical interpretation of Eq. (\ref{phi1})
then provides $\phi$ versus $X$ for a fixed $Z$, a result that is
plotted in Fig. \ref{comptheory}.  Using these two results the
entropy is obtained using Eq. (\ref{eqnofstate}).  Values of the
theoretical entropy density are plotted in Fig. \ref{enttheory}
for several values of $Z$.  By fixing $Z$, we are equivalently
imposing a fixed $\mu$ upon the system, as $Z(\mu)$ is determined
by $\mu$ exclusively within the confines of this model.

\begin{figure}
\centering { \vbox{
 \resizebox{8cm}{!}{\includegraphics{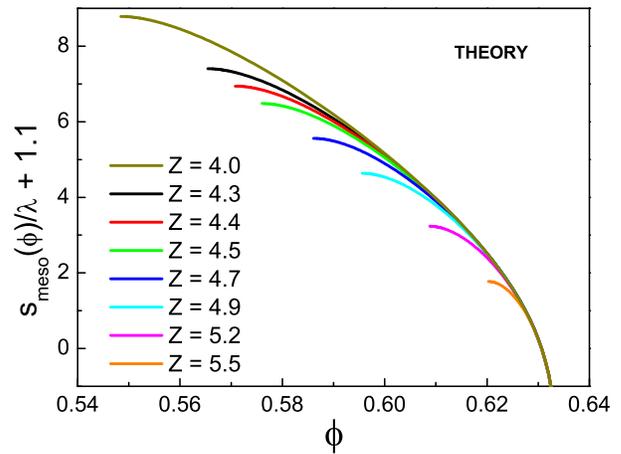}}
} } \caption{Prediction of the mesoscopic theory for $s_{\rm
meso}(\phi)$.} \label{enttheory}
\end{figure}

We see that the theoretical entropy density captures the general
behavior found in the simulations as shown in Fig.
\ref{entshift1}, i.e., it is maximal as we approach RLP for $Z=4$
and $X\to \infty$ while approaching the minimum entropy at RCP.
Furthermore, all the curves for different $Z$ approach $S\sim \ln
X$ as $X\to 0$, similar to a thermal ideal gas.  At the mesoscopic
level, the entropy vanishes at RCP.  In fact it diverges to
$-\infty$ when $\phi\to \phi_{\rm RCP}$ closer than a constant
proportional to $h_z$ (once again assuming an exponential
distribution for $g(z)$), much like the Planck constant imposes a
finite size in the phase space of quantum mechanics. Thus, we
assert that the value of $\phi$ at which $s(\phi) = 0$ in the
theoretical model provides a definition of RCP at the mesoscopic
level.  The theoretical $\phi(X)$ arising from Fig.
\ref{comptheory} is also in qualitative agreement with the
simulation results of Fig. \ref{compsim}.

We use $h_{z} = 0.01$ in Fig. \ref{enttheory} such that the
mesoscopic entropy vanishes very close to the predicted value of
$\phi_{\rm RCP} \approx 0.634$ \cite{jamming2}.  However, the
maximum value of $\phi_{\rm RCP} \approx 0.642$ from simulation,
introduces a discrepancy between the theoretical and simulated
models.

The value of $h_z$ is chosen to fit the mesoscopic theory of Fig.
\ref{enttheory} with simulation as close as possible, where the
only constraint imposed by theory is $h_z < 1$.  While the values
of both $\phi_{\rm RCP}$ and $\phi_{\rm RLP}(Z)$ from the theory
are well reproduced by the simulation results, it is clear that
the values of $s$ are not, as evidenced by comparing Fig.
\ref{enttheory} with Fig. \ref{entshift1}. For example, from
simulations we find $s_{\rm RLP} = 2.8\lambda$ at $Z = 4.3$ and
theory predicts $s_{\rm RLP} = 8.8\lambda$.  This is directly due
to the magnitude of $h_z$, and its implications towards the
density of states, $g(z)$. If we examine $s(X \rightarrow
\infty,Z)$, we achieve the entropy as a function of $Z$ along the
RLP line.  When $X \rightarrow \infty$, the equation of state in
Eq. (\ref{eqnofstate}) is rewritten as
\begin{equation}
s_{\rm meso}(X \rightarrow \infty,Z) = \lambda \ln
\int_{Z}^{Z_{max}} (h_z)^{z} dz,
\end{equation}
where $Z_{max} = 6.2$.  This equation is exactly solvable,
resulting in the following formula for the mesoscopic entropy
along the RLP line.

\begin{equation}
s_{\rm meso}(X \rightarrow \infty,Z) = \lambda \ln
\biggl(\frac{h_z^{Z_{max}} - h_z^{Z}}{\ln h_z} \biggr).
\label{srlp}
\end{equation}

Adjusting the value of $h_z$ directly affects the mesoscopic
entropy along the RLP as defined by Eq. (\ref{srlp}).  A similar
analysis has been used to obtain the functional form for $\phi(X
\rightarrow \infty,Z)$ along the RLP line.
\begin{eqnarray}
\begin{split}
&\phi(X \rightarrow \infty,Z) \\ =&\frac{\ln h_z}{h_z^{Z_{max}} -
h_z^{Z}} \int_{Z}^{Z_{max}} \frac{z}{z+2\sqrt{3}} (h_z)^z dz.
\label{phirlp}
\end{split}
\end{eqnarray}

Equation (\ref{phirlp}) is the equation of state for the RLP line,
and is plotted in Fig. \ref{phasesim} in the limit $h_z
\rightarrow 0$ when it reduces to Eq. (\ref{phirlpapprox}).
However, for a general $h_z$, Eq. (\ref{phirlp}) applies.   While
this equation is not exactly solvable, it is easily seen that
changing $h_z$ will not only impact the magnitude of the
mesoscopic entropy along the RLP line of Eq. (\ref{srlp}), but
also impact the values of $\phi_{\rm RLP}(Z)$ which define the
left most boundary of the phase diagram. Further, the effect on
one will be the inverse of the effect on the other. Simply stated,
within the present mesoscopic framework, one cannot satisfy
fitting both the value of the entropy at, for instance, RLP and
the value of $\phi_{\rm RLP}$ in the phase diagram from simulation
to the mesoscopic theory where $h_z$ is the only adjustable
parameter.

\subsection{Mesoscopic and Microscopic Fluctuations}

Near RCP the mesoscopic entropy vanishes. More specifically, it
diverges to $-\infty$ when $\phi\to \phi_{\rm RCP}$ closer than a
constant proportional to $h_z$, providing a characterization of
RCP at the mesoscopic level.  This result qualitatively resembles
behavior of the complexity of the jammed state in the replica
approach to jamming \cite{shannon,vink}.  We identify this point
as a mesoscopic ``Kauzmann point'', in analogy with the density,
or temperature, at which the configurational entropy of a
colloidal, or molecular, glass vanishes at the ideal glass
transition \cite{stillinger}. From this point the entropy
increases monotonically with $X$, being maximum for the RLP limit.
An important result is the direct implication of a larger number
of states available to jammed systems at RLP with respect to any
higher volume fraction, directly implying maximal entropy at the
RLP limit. Packings with packing fractions above RCP to $\phi_{\rm
FCC} = 0.74$, the optimal packing fraction for spheres in 3d, do
not appear in our theory because they exhibit some degree of
order, or crystallization.  By doing so, we explicitly do not
consider crystals or partially crystalline packings in the
ensemble. This is a direct consequence of setting the upper limit:
$z \leq 6$. States with $\phi > \phi_{\rm RCP}$ are new
microscopic states of the system, and their existence requires
further theoretical investigation.

At RCP, we find minimal fluctuation with respect to Voronoi
volumes associated to each grain.  This implies the surprising
conclusion of a minimal number of mesoscopic states for
frictionless systems at RCP.  Considering these minimal
fluctuations to be essentially zero, RCP has zero entropy and no
fluctuations with respect to a mesoscopic coarse-graining over the
ensemble. This is the frictionless jamming transition
\cite{makse,zhang} or J-point \cite{makse,zhang,ohern}.  We see
that, in principle, at the mesoscopic level this transition point
is well-defined. However a mesoscopic state parameterized by a
given average coordination number contains many microscopic states
which are averaged out in the coarse-graining procedure to
calculate the volume function at the quasi-particle level. For
instance, while the isostatic condition requires $Z = 2d$ contacts
per grain averaged over the entire packing, it makes no
implication towards the exact distribution of contacts per
individual grain. This allows for the existence of microscopic
states with grains having $Z < 2d$ and $Z
> 2d$ within the packing. Therefore we expect that these
microscopic states contribute to a nonzero entropy at the J-point
at RCP.  Indeed the Shannon entropy calculation of Section
\ref{simresults}.D finds the entropy of RCP to be $s_{\rm RCP} =
1.1\lambda$.

The full entropy should consider both mesoscopic and microscopic
contributions, such that
\begin{equation}
s = s_{\rm meso} + s_{\rm micro}. \label{sadd}
\end{equation}
The mesoscopic contribution is obtained via the theory, while the
microscopic contribution can be obtained herein using the Shannon
entropy method. We know that $s_{\rm RCP} = 1.1\lambda$ from
simulations and $s_{\rm meso} = 0$ from theory.  Therefore, the
total entropy at RCP is just the microscopic entropy, $s_{\rm RCP}
= s_{\rm micro}(RCP)$, which is equal to $1.1\lambda$ from
simulations.  Thus, $s_{\rm micro}(RCP) = 1.1\lambda$.  This
result is understood since all the jammed states are degenerate
around the mesoscopic ground state with the coordination number
$z=6$. As noted, these states still have slightly different volume
fractions, which leads to the microscopic fluctuations which are
coarse-grained in the mesoscopic theory. At the present time we do
not have a theory to explain the value of $1.1\lambda$ since the
mesoscopic theory does not include microscopic states nor
fluctuation in the coordination number. Next, we make the
additional assumption that the microscopic entropy is independent
of the volume fraction and can therefore consider $s_{\rm
micro}(\phi) = s_{\rm micro}(RCP) = 1.1\lambda$ for all values of
$\phi$ between RCP and RLP.  The result is the total entropy of
the packing as
\begin{equation}
s(\phi) = s_{\rm meso}(\phi) + 1.1\lambda \label{saddmicro}
\end{equation}
for any $\phi$, where $s_{\rm meso}$ is calculated by Eq.
(\ref{eqnofstate}). Equation (\ref{saddmicro}) is plotted in Fig.
\ref{enttheory}. Where the mesoscopic entropy is augmented by its
fixed value at RCP, and this value is $s_{RCP} = 1.1\lambda$. We
reiterate that in Fig. \ref{enttheory}, the addition of $1.1$ to
$s_{\rm meso}(\phi)$ makes two assumptions. First, its assumes
that $s_{\rm meso}(RCP) = 0$ as predicted by theory. Second, it
assumes that $s_{\rm micro}(\phi) = s_{\rm micro}(RCP) =
1.1\lambda$, as determined by the Shannon entropy method, and is
not an explicit function of $\phi$.

The mesoscopic entropy is $s_\mathrm{meso}=0$ not only at RCP but
for $\phi_{\rm RCP}\le\phi\le\phi_{\rm FCC}$.  This implies an
intrinsic difference between the current theory and Edwards'
statistics given by Eq. (\ref{exact}) leading to the separation of
the entropy in terms of the different length scales as in Eq.
(\ref{sadd}).

\subsection{Negative Compactivity}

Analyzing the partition function from a mathematical approach, we
are interested in the concept of negative compactivity and its
effect on the equation of state.

\begin{figure}
\centering { \vbox{
 \resizebox{8cm}{!}{\includegraphics{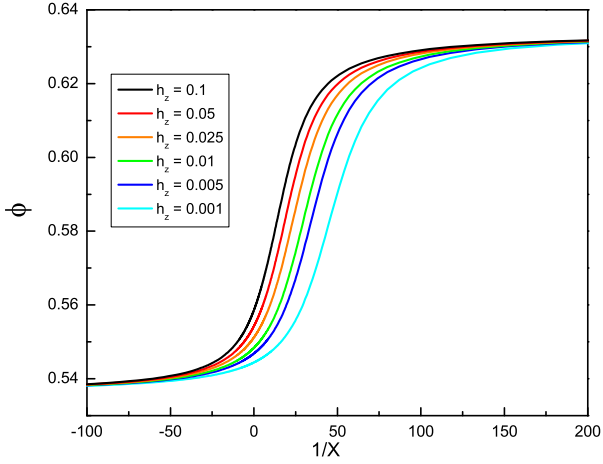}}
} } \caption{Volume fraction as a function of the inverse of
compactivity, for Z = 4, as calculated using theoretical methods,
including negative compactivity values, for $h_z$ ranging from
$0.001$ to $0.1$. When $X$ discontinuously jumps from $\infty$ to
$-\infty$, $\phi(X)$ exhibits continuous behavior. However, when
$X$ continuously goes from $0^-$ to $0^+$, $\phi(X)$ exhibits a
discontinuous jump from $\approx 0.536$ to $\approx 0.634$. }
\label{negcomp}
\end{figure}

\begin{figure}
\centering { \vbox{
 \resizebox{8cm}{!}{\includegraphics{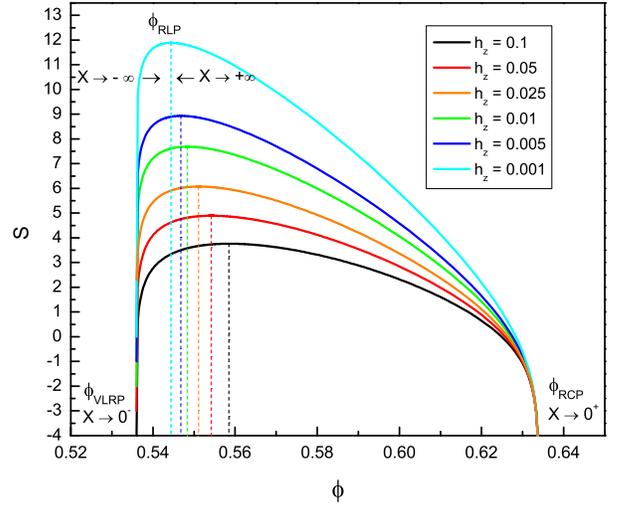}}
} } \caption{Entropy plotted for Z = 4, as calculated using
theoretical methods, including negative compactivity values.
$S(\phi)$ tends towards $-\infty$ at $\phi = 0.536$ ($X
\rightarrow 0^-$) and $\phi = 0.634$ ($X \rightarrow 0^+$), the
minimal value of RLP and RCP, respectively, for any value of
$h_z$. Dashed vertical lines show $\phi(X)$ for $X \rightarrow \pm
\infty$, acknowledging the lowest physically achievable volume
fraction at $Z = 4$ for a particular value of $h_z$.  Larger
values of $h_z$ result in larger values of $\phi(X \rightarrow \pm
\infty)$.  Following \cite{rvlp} we call $\phi(X \rightarrow 0^-)
\rightarrow \phi_{\rm VLRP}$ (very loose random packing), $\phi(X
\rightarrow \pm \infty) \rightarrow \phi_{\rm RLP}$ and $\phi(X
\rightarrow 0^+) \rightarrow \phi_{\rm RCP}$.  In the limit of
$h_z \rightarrow 0$ the difference between VLRP and RLP vanishes
$\Rightarrow {\phi_{\rm VLRP} \atop {h_z \rightarrow 0}}
\rightarrow  \phi_{\rm RLP}$.  The maximum entropy is always at
RLP. Negative compactivity states are very difficult to obtain
with current protocols. } \label{negent}
\end{figure}

When $X \rightarrow 0^{+}$, $\phi(X \rightarrow 0^{+}) \rightarrow
\phi_{\rm RCP}$. Under the assumption of a very large $g(z)$ at
$z=Z$, with respect to any higher $z$, $X \rightarrow +\infty$,
$\phi(X \rightarrow +\infty) \rightarrow \phi_{\rm RLP}(Z) =
\frac{Z}{Z+2\sqrt{3}}$, where $Z$ is the lower limit of
integration in Eq. (\ref{phi1}).  This occurs because when using
the density of states of Eq. (\ref{gz}), with $h_z \rightarrow 0$,
$g(z) \sim \delta(z - Z)$.  However, as evidenced above, an
exponential form for $g(z)$ may not well reproduce simulation
results for the entropy. Altering $g(z)$ is shown to better
reproduce the entropy equation of state, but shifts the predicted
value of $\phi_{\rm RLP}$ higher. Examination of a negative
compactivity within the above presented statistical mechanics
framework should allow us to achieve the minimum value of RLP in
the limit $X \rightarrow 0^-$.  For a background on negative
temperature states in equilibrium statistical mechanics see Landau
and Lifshitz book on statistical physics.  Note that a negative
temperature state is "hotter" than a state at absolute zero
temperature and any positive temperature state.  The state with $T
\rightarrow +\infty$ is physically identical to the state with $T
\rightarrow -\infty$.  As in magnetic systems where negative
temperature states can be observed, granular matter is
characterized by a bounded volume fraction and "thermalization" of
volume at a negative temperature is possible, in principle.  Next,
we analyze those states with a negative compactivity.

When $X \rightarrow 0^{+}$, the Boltzmann factor in the partition
function of Eq. (\ref{pf1}) tends towards zero. As such, the
largest value of $z$, or the smallest value of $1/z$ will give the
largest value of the Boltzmann factor when using the partition
function to calculate observable averages.  This results in the
calculation of the RCP state.  However, when $X \rightarrow
0^{-}$, the Boltzmann factor in the partition function tends
towards infinity, not zero, and the largest value of $Z$, or the
smallest value of $1/z$ will give the largest value of the
Boltzmann factor.  When calculating the average volume function in
either case, the density of states will not greatly impact the
results with respect to the contribution from the Boltzmann
factor. Therefore, when $X \rightarrow 0^{-}$, the average volume
fraction will reduce to the predicted value of the RLP line where
$h_z \rightarrow 0$, $\phi(X \rightarrow 0^{-}, Z) = \phi_{\rm
RLP}(Z) = \frac{Z}{Z+2\sqrt{3}}$. Figure \ref{negcomp} exemplifies
this phenomenon, plotting $\phi(X)$ at $Z = 4$ for different
values of $h_z$.

The entropy equation of state should achieve the same values as $X
\rightarrow +\infty$ and $X \rightarrow -\infty$, since the
Boltzmann factor approaches unity in either case.  Although not as
obvious, the same can be said for $X \rightarrow 0^{+}$ and $X
\rightarrow 0^{-}$.  The equation of state (\ref{eqnofstate}) can
be rewritten as:

\begin{equation}
s_{\rm meso}(X,Z)/\lambda = \ln \int_{Z}^6 (h_z)^{z-2d}
e^{-\frac{1}{\lambda X} (w(z) - \langle w(z) \rangle)} dz.
\label{entrewrite}
\end{equation}

In the case of $X \rightarrow 0^{+}$, $X$ is positive, and $w(z)
\geq \langle w(z) \rangle$, as RCP represents the lowest
attainable value of $\langle w(z) \rangle$.  The exponentiated
term in the partition function is always negative, such that the
entropy approaches $-\infty$.  Conversely, when $X \rightarrow
0^{-}$, $X$ is negative, and $w(z) \leq \langle w(z) \rangle$, and
we obtain the highest attainable value of $\langle w(z) \rangle$.
Again, the exponentiated term in the partition function is always
negative, such that the entropy approaches $-\infty$.

Figure \ref{negent} displays the equation of state for $Z = 4$ as
calculated by the theoretical model of Eq. (\ref{entrewrite}). We
use $Z = 4$ and $Z_{max} = 6$ as the limits of integration for the
partition function of Eq. (\ref{pf1}). The value of $h_z$ is
increased to show that larger values of $g(z)$ at RLP result in a
more precise agreement between $\phi(X \rightarrow \pm \infty)$
and $\phi(X \rightarrow 0^-)$. The dashed vertical lines of Fig.
\ref{negent} show $\phi(X)$ for $X \rightarrow \pm \infty$,
acknowledging the lowest physically achievable volume fraction at
$Z = 4$ for a particular value of $h_z$.  Larger values of $h_z$
result in larger values of $\phi(X \rightarrow \pm \infty)$,
allowing for the existence of a greater range of packings with a
negative compactivity, having $\phi(X \rightarrow 0^{-}) < \phi
<\phi(X \rightarrow -\infty)$.

Some models exist where the concept of a negative temperature
finds physical meaning, including nuclear spins and semiconducting
lasers.  In Ref. \cite{rvlp} an attempt is made to include the
concept of a negative temperature within a statistical mechanics
framework.  By means of a lattice model in $2d$, utilizing a
discrete phase space, mechanically stable packings, or
microstates, are shown to exist beneath the volume fraction with
the largest number of microstates, at a particular $\mu$.  The
highest entropy occurs when the largest number of microstates are
available, and is the equivalent of $\phi_{\rm RLP}$ in the above
presented mesoscopic theory.  Under the assumption of an ergodic
exploration of the volume fractions available to the lattice
model, this implies that packings with $\phi < \phi_{\rm RLP}$
exist, with entropy below the maximal value, and that these
packings can be explained via the concept of a negative
temperature.  Reference \cite{rvlp} thereby considers $\phi_{\rm
RLP}$ to be the "loosest possible random packing that is
mechanically stable that one can achieve by pouring grains". Below
this limit there exists RVLP "random very loose packings" with
negative temperature.

Although the present work analyzes packings in $3d$, the results
of mesoscopic theory find agreement with the simulations of
\cite{rvlp}.  The main results of \cite{jamming2}, Fig.
\ref{phasesim} and for instance Eq. (\ref{phirlpapprox}) have been
obtained in the limit of $h_z \rightarrow 0$.  If we relax this
constraint then negative compactivity states are possible.  We
assert that these states exist because there is an upper bound in
the volume function at $z = Z$.  Systems with unbounded volume
functions (Hamiltonians) do not allow for negative compactivity
(temperature) states.  In this context we have the definition of
the following limits according to the entropy and compactivity:
The RCP limit is $\phi(X \rightarrow 0^+)  = \phi_{\rm RCP}$ and
minimum entropy: Neglecting the negative diverging of the entropy
we have $S_{\rm RCP} = S(X \rightarrow 0^+) \rightarrow 0$.  The
RLP limit is defined as the maximum entropy in the limit $\phi(X
\rightarrow \pm \infty)  = \phi_{\rm RLP}$.  If $h_z$ is finite
then VLRP appears as $\phi(X \rightarrow 0^-) = \phi_{\rm VLRP}$
and minimum entropy.  Again, neglecting the divergency, we obtain
$S_{\rm VLRP} = S(X \rightarrow 0^-) \rightarrow 0$.  In the limit
of $h_z \rightarrow 0$ the difference between RLP and VLRP
vanishes and we have only one well defined RLP as: ${\phi_{\rm
RLP} \atop {h_z \rightarrow 0}} \rightarrow \phi_{\rm VLRP}$.

Fig. \ref{negent} shows a maximal entropy at $\phi_{\rm RLP}$,
indicating the largest number of available microstates to the
system.  The introduction of a negative compactivity, as described
above, allows the theory to probe states such that $\phi <
\phi_{\rm RLP}$.  It becomes apparent that the range of $\phi$ in
which these states may exist is directly related to the magnitude
of $h_z$, decreasing as the discretization of phase space within
the confines of the mesoscopic theory such that in the limit of a
continuous phase space none of these packings are mechanically
stable.  Our theoretical model includes the concept of a discrete
phase space for jammed grains, and our simulations show that $h_z
< 1$, but not necessarily $h_z \ll 1$.  Further, the "split"
algorithm utilized simulates a pouring of grains with respect to
the method of packing creation.  It is possible that the concept
of a negative compactivity could help to explain areas in the
phase diagram of Fig. \ref{phasesim} unavailable within the scope
of the present study.

\section{Outlook}
\label{outlook}

Although extensive detail is presented in this study regarding the
various steps necessary to analyze the equations of state, several
questions remain unclear.

\textbf{(a)} The derivation of the entropy for jammed granular
matter is explicitly calculated via fluctuations in Voronoi cell
volume for each packing presented herein.  As we are studying
random packings, crystal states are not achieved with anything
other than measure zero probability.  Thus, the highest available
volume fraction for any given packing is RCP, $\phi \simeq 0.64$,
not FCC, $\phi = 0.74$, as corroborated by simulation results.
This result is the apparent limit of the preparation protocols
used herein.  It can be said with certainty that a FCC packing has
zero fluctuation with respect to their constituent Voronoi cells.
At RCP simulations reveal a non-zero fluctuation, as discussed
above. This approach, however, does not account for the
possibility that for packings above RCP, but below FCC, may
achieve a continuous, monotonic, degree of crystallization. Such a
condition would permit a continuous decrease in fluctuation to
exactly zero, from RCP to FCC, which can be studied within the
scope of a more complete theory that includes random and partially
crystallized packings.

Below we elaborate on a possible scenario to rationalize the
transition from disorder at RCP to order at FCC.  If the
microscopic fluctuations are not subtracted from RCP then the
compactivity curves presented in Fig. \ref{compsim} no longer
reach a plateau when approaching RCP, but reach a finite,
non-zero, value.  This opens the possibility that a true
thermodynamic phase transition may occur at RCP between a
disordered phase and an ordered phase.  It remains possible that a
phase transition occurs at RCP, and packings of higher volume
fraction need not preserve the properties of a fully random
system.  It remains an open topic how one would define
compactivity between RCP and FCC, as compactivity as been herein
attributed to packing protocols resulting in random packings.

The mesoscopic theory of \cite{jamming2} utilized herein considers
a vanishing entropy at, or near, RCP.  Taking into consideration
the FCC state, a more complete theory including packings between
RCP and FCC should be characterized, such that the entropy of
jammed matter approaches zero when approaching FCC, not RCP.
Furthermore, the result that $X\rightarrow0$ at RCP is merely an
artifact of the mesoscopic theory that neglects microscopic
fluctuations.  A full theory would obtain $X\rightarrow0$ at FCC.
Figure \ref{rcpfcc} displays a possible interpretation for an
extension of the entropic equation of state. The entropy attains a
maximal value at $\phi_{\rm RLP}$, as predicted by the existing
mesoscopic theory. When $\phi_{\rm RLP} < \phi < \phi_{\rm RCP}$,
the packing consists of purely random states, and the entropy
decreases as we approach RCP from lower volume fractions, also as
predicted. At some point close to $\phi_{\rm RCP}$, the entropy
deviates from its predicted decrease to zero at $\phi_{\rm RLP}$,
and follows a different branch.  When $\phi_{\rm RCP} < \phi <
\phi_{\rm FCC}$ a coexistence between random and crystalized
microstates may exist, ultimately leading to a purely crystalized
packing at FCC. The exact nature of the transition, continuous or
discontinuous, from purely random states to a coexistence of
states remains an open topic. The incorporation of microscopic
fluctuations and microscopic crystalized states into the existing
mesoscopic theory may result in a more complete characterization
of the entropy of jammed granular matter, a work currently in
progress.

\begin{figure}
\centering { \vbox{
 \resizebox{8cm}{!}{\includegraphics{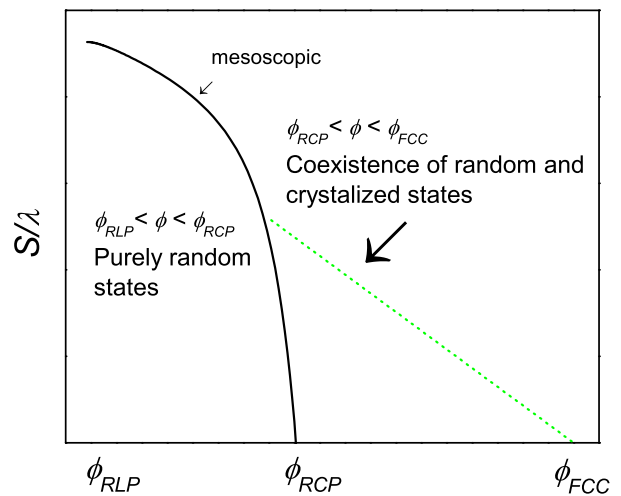}}
} } \caption{A possible extension of the entropy equation of state
to include packings between RCP and FCC. For $\phi_{\rm RLP} <
\phi < \phi_{\rm RCP}$, the packing consists of only random
states, and the entropy decreases when approaching RCP from lower
volume fractions, as predicted by the mesoscopic curve.  At
$\phi_{\rm RCP}$, the entropy does not decrease to zero, as
predicted by the mesoscopic curve, and follows a different branch,
achieving exactly zero entropy at $\phi_{\rm FCC}$. When
$\phi_{\rm RCP} < \phi < \phi_{\rm FCC}$ a coexistence between
random and crystalized microstates may exist.  The possible
existence of a transition that may define RCP remains an open
question. } \label{rcpfcc}
\end{figure}

\textbf{(b)} An additional assumption made in calculating the
compactivity of jammed granular matter is that those packings
along the RLP line have $X = \infty$.  This assumption directly
allows for the $\phi(X)$ equation of state without any constants
due to integration techniques.  It appears to be reasonable to
presume that $X$ is large along the simulated RLP line, with
respect to $X$ of all other packings used herein, and that the
constant term $\frac{1}{X(\phi_{\rm RLP})}$ is very small.

\textbf{(c)} As mentioned above, there exists a small fluctuation
density in Voronoi volume at RCP.  The fluctuation density
increases as we increase $n$, and comes to a maximal value at a
particular range of $n$, where $n$ is larger as we approach lower
values of $\phi$ as shown in Fig. \ref{extensive}.  This increase
in value is due to selecting of clusters of $n$ Voronoi volumes,
as opposed to $n$ randomly chosen grains.  In the case of randomly
chosen grains, the fluctuation density remains constant, equal to
the fluctuation at $n = 1$.  Such a change implies a correlation
between the Voronoi volumes in clusters of grains, exemplified by
Eq. (\ref{corr1}).  This correlation may create a scale
separation, such that microscopic correlations dominate the
fluctuations at lower $n$ and mesoscopic correlations dominate at
higher $n$. This scale separation would result in a difference
between local and global compactivities, suggested in
\cite{aste2}, and is a topic of continuing study.  It also remains
possible that fluctuation densities calculated using larger $n$
sized clusters overuse the data set by largely repeating
constituent Voronoi cells in cluster volumes, thereby rendering
the fluctuation densities questionable for packings approaching
RLP.  Again, it should be noted that simply using the microscopic
fluctuations does not greatly effect the entropy calculations.

\textbf{(d)} The mesoscopic theory predicts $S \rightarrow
-\infty$ as we approach RCP for any $\mu$.  This concept does not
adhere to physical measurements, where the minimal entropy should
be zero. Taking this into consideration, another method must be
used to calculate the entropy of packings at RCP, independent of
the distribution of Voronoi volumes used to facilitate the
mesoscopic theory.  The Shannon entropy calculation applies graph
theoretical methods, resulting in a non-zero value for the entropy
of RCP, well suited for a more complete equation of state.  While
it can be said with some degree of certainty that the Shannon
entropy calculation contributes to the entropy at RCP, it is
unclear whether it does so completely.  Other methods may be
available that provide all of the entropy at RCP, or give
additional terms to the Shannon entropy, another topic of
continuing study. Further, as discussed above with respect to
fluctuation density, packings between RCP and FCC have partial
degrees of crystallization, such that the entropy of an FCC is
exactly zero, or $-\infty$ as described in the theory.  Recent
work \cite{aste3} suggests that the entropy experiences an
increase immediately following RCP, due to mixing between random
and crystal states. Incorporation of these ideas into the present
work remains a topic of ongoing study.

\textbf{(e)} While the concept of negative compactivity works well
as a mathematical tool for achieved the VLRP state of jammed
granular matter, it remains difficult to attribute a physical
meaning to such a condition.  Recent studies in \cite{rvlp}
suggest that negative compactivity probes mechanically stable
states that exist beneath RLP, but are not accessible by means of
grain pouring. It should be noted that $X<0$ states may not be
thermalized with $X>0$.  Such ideas will remain the topic of
future studies.

\textbf{(f)} An exponential density of states with a very small
value of $h_z$ provides a very accurate phase diagram, in
comparison with simulations, but values for the entropy that are
different by a factor of $4$ times larger than those achieved via
simulation.  It remains possible that the methods presented herein
simply do not capture all of the entropy for jammed matter, and
larger values from simulation are possible, such that corrections
to the mesoscopic theory could be minimized. Further, a new theory
that carefully analyzes all packings up from $\phi_{\rm RLP}$
through $\phi_{\rm FCC} = 0.74$ may result in the ability to
capture all of the entropy for a given packing. Current work is
approaching this idea from a microscopic level.

\textbf{(g)} The differences displayed between theoretical and
simulated entropy can to some extent be considered within the
scope of disagreement between classical and quantum entropy.
Classical entropy measurements are primarily interested in $\Delta
S$, having a minimum value of $0$ when approaching the ground
state, though $S \rightarrow -\infty$.  Quantum entropy
measurements assume a minimum value of phase space over which one
can integrate degrees of freedom for a give system.  This results
in a $S = 0$ exactly at the ground state.  Determination of a well
defined minimum phase space volume for jammed matter would adjust
the number of available microstates, $\Omega$, within the
micro-canonical ensemble as presented by Edwards.  Such a model
may be available through a careful analysis of the microscopic
entropy, including partially crystalized states above $\phi_{\rm
RCP}$ through the crystal state of $\phi_{\rm FCC}$, the highest
achievable packing fraction for identical spheres in $3d$.

\section{Conclusions}

Simulation results as derived from mesoscopic fluctuations are
presented in Figs. \ref{compsim} and \ref{entsim}.  When comparing
all the packings with different $Z(\mu)$ and $\phi$, the maximum
entropy is at the minimum volume fraction of RLP $\phi_{\rm RLP}$
when $X\to\infty$ but only infinite friction. The minimum entropy
is found for the RCP state at $\phi_{\rm RCP}$ for $X\to 0$, now
for all the values of friction, indicating the degeneracy of the
RCP state. It is commonly believed that the RCP limit corresponds
to a state with the highest number of configurations and therefore
the highest entropy.  This belief is expressed for instance in the
definition of RCP as the maximally random jammed state
\cite{salvatore1}. However, here we show that the states with a
higher compactivity have a higher entropy, corresponding to the
looser RLP packings.  Within a statistical mechanics framework of
jammed matter, this result is a natural consequence, and gives
support to such an underlying statistical picture.

Each curve in the Figs. \ref{comptheory} and \ref{enttheory}
correspond to a system with a different $Z(\mu)$, as calculated
using a mesoscopic ensemble. When comparing all the packings, the
maximum entropy is at $\phi_{\rm RLP}$ and $X\to \infty$ while the
entropy is minimum for $\phi_{\rm RCP}$ at $X\to 0^+$. Following
the $Z = 4$ line in the phase diagram we obtain the entropy for
infinitely rough spheres showing a larger entropy for the RLP than
the RCP. The same conclusion is obtained for the other packings at
finite friction ($4<Z(\mu)<6$). We conclude that the RLP states
are more disordered than the RCP states. Approaching the
frictionless $J$-point, $\mu \to 0$ ($Z=6$) the entropy vanishes.
More precisely, it vanishes for a slightly smaller $\phi$ than
$\phi_{\rm RCP}$ of the order $h_{z}$. Strictly speaking it
diverges to $-\infty$ at $\phi_{\rm RCP}$ as $S \rightarrow \ln X$
for any value of $Z$, in analogy with the classical equation of
state. However, this is an unphysical limit, as it would be
considering distances in phase space smaller than the minimal
distance in the jamming phase space.  Thus we consider only
packings with an entropy density greater than or equal to $0$ as
"physical" packings.  We note that the compactivity curves from
the theoretical model match simulation with accuracy. The
theoretical entropy fails to agree with the entropy from
simulation in magnitude, but reproduces the overall shape.  While
increasing $h_z$ such that the magnitude of the entropy from the
mesoscopic theory would decrease, the RLP line would no longer be
well reproduced.  As simulations and theory are in strong
agreement with respect to the RLP line, increasing $h_z$ does not
appear to be a reasonable amendment to the mesoscopic partition
function.

In summary, a notion of disorder is presented that applies to
frictional hard spheres, as well as frictionless ones.  The
entropy reveals interesting features of the RCP and RLP states
such as the fact that RLP is maximally random with respect to RCP
and that both limits can be defined in terms of the entropy and
equation of state.  Overall, the agreement between theory and
simulation is sufficient to indicate that the methods presented
herein are appropriate for evaluating the entropy of jammed
matter.

Acknowledgements. - We express our thanks for the financial
support of NSF and DOE. We further thank Kun Wang and Yuliang Jin
for insightful discussions.


\begin{thebibliography}{99}


\bibitem{edwards} S. F. Edwards and D. V. Grinev,
in Jamming and Rheology, Eds. A. Liu and S. R. Nagel (Taylor and
Francis, New York, 2001).

\bibitem{coniglio} A. Coniglio, A. Fiero, H. J. Herrmann and M. Nicodemi eds.
{\it{Unifying Concepts in Granular Media and Glasses}}
(Elsevier, Amsterdam, 2004).

\bibitem{liu-nagel} A. J. Liu and S. R. Nagel,
Nature {\bf 396}, 21 (1998).

\bibitem{makse} H. A. Makse, D. L. Johnson, and L. M. Schwartz,
Phys. Rev.  Lett. {\bf 84}, 4160 (2000);

\bibitem{zhang}
H. Zhang and H. A. Makse, Phys. Rev. E {\bf 72} 011301 (2005).

\bibitem{ball} R. C. Ball and R. Blumenfeld, Phys. Rev. Lett. {\bf
88}, 115505 (2002).

\bibitem{ohern} C. S. O'Hern, S. A. Langer, A. J. Liu, S. R. Nagel,
Phys. Rev. Lett. {\bf 88}, 075507 (2002).

\bibitem{jamming2} C. Song, P. Wang, H. A. Makse, Nature {\bf 453}, 629 (2008):arXiv:0808.2196.

\bibitem{onoda}
Onoda, G. Y. \& Liniger, E. G. Phys. Rev. Lett. {\bf 64},
2727-2730 (1990).

\bibitem{bernal} J. D. Bernal, Nature {\bf 185}, 68 (1960).

\bibitem{salvatore2} S. Torquato, T. M. Truskett, and
P. G. Debenedetti, Phys. Rev. Lett. {\bf 84}, 2064 (2000).

\bibitem{salvatore1} S. Torquato and F. H. Stillinger, J. Phys. Chem B
{\bf 105}, 11849 (2001).

\bibitem{kertesz}
T. Under, J. Kertesz, and D. E. Wolf, Phys. Rev. Lett. {\bf 94},
178001 (2005)

\bibitem{sirsam} S. F. Edwards and R. B. S. Oakeshott,
Physics A {\bf 157} 1080 (1989).

\bibitem{alexander} S. Alexander, Phys. Rep. {\bf 296}, 65 (1998).

\bibitem{silbert}
L. E. Silbert, {\it et al.} Phys. Rev. E {\bf 65}, 031304 (2002).

\bibitem{stealing} Stealing the gold: a celebration of the pioneering
physics of Sam Edwards, P. M. Goldbart, N. Goldenfeld, D.
Sherrington, eds.  (Oxford Science Publications, Oxford, 2004).

\bibitem{moukarzel} C. F. Moukarzel,
Phys. Rev. Lett {\bf 81}, 1634 (1998)

\bibitem{vanvan} W.G. Ellenbroek, E. Somfai, M. van Hecke, and W. van
Saarloos, Phys. Rev. Lett. {\bf 97}, 258001 (2006).

\bibitem{rvlp}
M. P. Ciamarra and A. Coniglio, Phys. Rev. Lett. {\bf 101}, 128001
(2008).

\bibitem{landau} L. D. Landau and E. M. Lifshitz, Theory of
Elasticity, (Pergamon, NY, 1970).

\bibitem{mindlin} R. D. Mindlin, J. Appl. Mech. (ASME) {\bf 71}, (1949).

\bibitem{jamming1} C. Song, P. Wang, H. A. Makse: arXiv:0808.2186.

\bibitem{Hales} T.C. Hales, The Kepler Conjecture, http://arxiv.org/abs/math.mg/9811078

\bibitem{chicago2} E. R. Nowak, J. B. Knight, E. BenNaim, H. M. Jaeger
and S. R.  Nagel, Phys. Rev. E {\bf 57}, 1971 (1998).

\bibitem{swinney} M. Schr$\ddot{\rm o}$ter, D. I. Goldman, H. L. Swinney,
Phys. Rev. E {\bf 71}, 030301(R) (2005).

\bibitem{dauchot} F. L\'echenault, O. Dauchot and E. Bertin,
J. Stat. Mech., P07009 (2006).

\bibitem{kumar2} V. S. Kumar and V. Kumaran, Journal of Chemical Physics {\bf 123}, 074502
(2005).

\bibitem{aste3}
T. Aste, A. V. Anikeenko and N. N. Medvedev, Phys. Rev E {\bf 77},
031101 (2008)

\bibitem{shannon} C. E. Shannon, Bell Sys. Tech. J. {\bf{27}}, 379 (1948).

\bibitem{vink} R. L. C. Vink and G. T. Barkema, Phys. Rev. Lett. {\bf 89},
076405 (2002).

\bibitem{kumar1} V. S. Kumar and V. Kumaran, Journal of Chemical Physics {\bf 123}, 114501
(2005).

\bibitem{brujic}
J. Bruji\'c, C. Song, P.Wang, C. Briscoe, G. Marty, and H. A.
Makse, Phys. Rev. Lett. {\bf 98}, 248001 (2007).

\bibitem{mckay} B. D. McKay, Nauty user's guide (version 1.5),
Tech. Rep. TR-CS-90-02, Australian National University (1990).

\bibitem{rombouts} P. Rombouts, Masters Thesis,  Institute for Theoretical Physics, Utrecht University, (2004)

\bibitem{parisi} G. Parisi and Zamponi, J. Chem. Phys. {\bf 123},
144501 (2005), arXiv:0802.2180

\bibitem{bruijicwang} J. Bruji\'c, P. Wang, D. Johnson, O. Sindt, and H. A.
Makse, Phys. Rev. Lett. {\bf 95}, 128001 (2005).

\bibitem{maksekurchan} H. A. Makse and J. Kurchan, Nature {\bf 415}, 614
(2002).

\bibitem{blum1} S.F. Edwards, Physica A, {\bf 353}, 114 (2005).

\bibitem{blumed} R. Blumenfeld, On Entropic Characterization of Granular Materials in Lecture Notes in Complex Syetems Vol 8:
Granular and Complex Materials, {\bf 43-53} (2007)

\bibitem{chak} S. Henkes, B. Chakraborty, Phys. Rev. Lett. {\bf 95}, 198002 (2005).

\bibitem{serb} S. Ostojic, E. Somfai, B. Nienhuis, Nature {\bf 439}, 828 (2006).

\bibitem{vanhecke} J. H. Snoeijer, T. J. H. Vlugt, M. van Hecke, W. van Saarloos, Phys. Rev. Lett. {\bf 92},
054302 (2004).

\bibitem{dauchot1}
E. Bertin, O. Dauchot, and M. Droz, Phys. Rev. Lett. {\bf 93},
230601 (2004).

\bibitem{landau-stat-mech} L. D. Landau and E. M. Lifshitz, Statistical Physics (Pergamon, NY, 1970).

\bibitem{stillinger}
F. H Stillinger, Science {\bf 267}, 1935-1939 (1995).

\bibitem{aste2}
T. Aste and T. Di Matteo, Phys. Rev. E {\bf 77}, 021309 (2008)

\end{thebibliography}
\end{document}